\documentstyle[12pt,aaspp4]{article}

\def\insertplot#1#2#3#4#5#6#7{
\vskip 10pt\nobreak\hbox to \hsize{\hss\dimen0=#3in\hbox to #6\dimen0{%
\dimen0=#2in\vbox to #6\dimen0{\vss
\special{ps: plotfile #1}
\special{ps::[end]
  PGPLOT restore
}
}\hss}\hss}\vskip 10pt}

\lefthead{Meyer {\it et al.}}
\righthead{Near--IR Spectroscopy}
\begin{document}

\title{Near--Infrared Classification Spectroscopy:  \\
H--band Spectra of Fundamental MK Standards}

\author{Michael R. Meyer\altaffilmark{1}} 
\affil{Five College Astronomy Department\\ University of
Massachusetts\\ Amherst, MA 01003\\ mmeyer@as.arizona.edu}

\author{Suzan Edwards}
\affil{Five College Astronomy Department\\ Smith College, Northampton, MA 01063 
\\ edwards@makapuu.ast.smith.edu}

\author{Kenneth H. Hinkle}
\affil{Kitt Peak National Observatory\altaffilmark{2}
\\National Optical Astronomy Observatories,
Tucson, AZ 85721 
\\ hinkle@noao.edu}

\author{Stephen E. Strom}
\affil{Five College Astronomy Department\\ University of
Massachusetts, Amherst, MA 01003 
\\ sstrom@tsaile.phast.umass.edu}

\altaffiltext{1}{Current Address: Hubble Fellow, 
Steward Observatory, The University of Arizona, 
933 N. Cherry Ave., Tucson, AZ 85721--0065}
\altaffiltext{2}{Operated by the Association of Universities for Research 
in Astronomy, Inc. under cooperative agreement with the National Science Foundation.}

\vfill\eject

\begin{abstract}

We present a catalogue of H--band spectra for 85 stars of approximately solar 
abundance observed at a resolving power of 3000 with the KPNO Mayall 4m FTS.  
The atlas covers spectral types O7--M5
and luminosity classes I-V as defined on the MK
system.  We identify both atomic and molecular indices 
and line--ratios which are temperature and luminosity
sensitive allowing spectral classification to be carried
out in the H--band.  The line ratios permit spectral classification 
in the presence of continuum excess emission, which is commonly found
in pre--main sequence and evolved stars.  We demonstrate that
with spectra of $R = 1000$ obtained at $SNR > 50$ it is 
possible to derive spectral types within $\pm 2$ subclasses 
for late--type stars. 
These data are available electronically through the Astronomical 
Data Center 
in addition to being served on the World--Wide--Web.

\end{abstract}


\section{Introduction}

With the recent development of large--format infrared array detectors, 
high quality photometric surveys are routinely conducted at wavelengths
between 1--2.5 $\mu$m.  Soon the completion of the 
2 Micron All--Sky Survey (2MASS; Skrutskie 1997) and
DENIS (Epchtein et al. 1997) will provide comprehensive catalogues of 
near--infrared sources with detection limits sensitive to a wide 
variety of stellar and non--stellar objects.  Infrared spectra will be 
required for appropriate identification of many of these sources, 
and for further study of their astrophysical properties. 

The pioneering study of Johnson and Mendez (1970) was the first to explore 
the spectra of a large sample of normal stars in the near--infrared.   
However many years 
passed before improvements in instrumentation made possible similar 
observations of large numbers of targets of astrophysical interest. 
The majority of the work done in near--infrared spectroscopy to date has
been focused on the K--band, in large part because intrinsically
cool or heavily obscured objects are typically brighter at K--band than 
in the J-- or H--bands.  
In 1986, Kleinmann and Hall (1986; KH86) provided the
first comprehensive medium resolution atlas ($R=3000$)
of stellar spectra in the K-band, covering
all luminosity classes, but restricted to spectral types between
F8-M7. More recently, Wallace and Hinkle (1997; WH97) have 
extended the KH96 K--band atlas, using the 
same FTS spectrograph on the KPNO 4m with
$R=3000$, but including stellar spectra spanning spectral types O-M and
luminosity classes I-V. They also summarize 
the considerable body of work directed toward K--band spectroscopy in the
last decade.

While in many situations, the K--band will be the wavelength
selection of choice for spectroscopic studies of highly 
obscured or very cool objects, the presence of 
circumstellar dust ($T_{vap} < 2000 K$; Pollack et al. 1994) 
often results in significant
excess continuum emission longward of $2~\mu$m. This continuum excess is
commonly found in two important classes of 
objects:  young stars with circumstellar disks 
(e.g. Meyer, Calvet, \& Hillenbrand, 1997) 
and evolved stars with extensive envelopes from mass-loss 
(e.g. Le Bertre 1997).  Near--infrared excess due to 
warm dust can also complicate spectroscopic studies 
of composite stellar systems aimed at discerning the stellar
populations of other galaxies (e.g. Schinnerer et al. 1997).
Continuum excess longward of $2~\mu$m will weaken or even render 
invisible the photospheric features in the K--band, while
the photosphere will dominate at shorter wavelengths. In such a
situation, near infrared spectra shortward of 2 $\mu$m will
be required to see the stellar photosphere too obscured to be
detected optically. 
To date, there has been relatively little work in the H--band (1.55--1.75 
$\mu$m). Recent publications include:  
i) observations of 40 G, K, and M stars of luminosity class I and III 
at $R=1500$ by Origlia, Moorwood, and Oliva (1993); ii) the library of 
56 spectra O--M of luminosity class I, II, and V at $R=500$ 
(Lancon \& Rocca--Volmerange, 1992); 
iii) a library of 37 stars of luminosity classes I, III, V
at $R=1500-2000$ (Dallier, Boisson, \& Joly 1996) over a limited portion of the
H--band; and iv) a study of 9 OB Stars at $R=570$ (Blum et al. 1997).  

Here we present an H--band spectral atlas 
for 85 stars of nearly solar abundance with spectral types on the
MK classification system ranging from 07-M5 and luminosity classes
I-V. These $R=3000$ spectra were collected with the same FTS at the KPNO 4m
as the K--band atlases of KH86 and WH97. 
In Section 2, we describe the sample selection and 
in Section 3 we describe the observations and calibration of the data. 
In Section 4 we discuss the dependence of the spectral features on
temperature and luminosity and suggest a two--dimensional 
classification appropriate for late-type stars.
In Section 5 we discuss near--IR spectral classification with regard to 
wavelength range/spectral resolution, and conclude with a summary of 
our results. 

\section{Defining the Sample}

In our sample selection, we chose optically visible stars which
had previously been identified on the temperature and luminosity
scales of the revised MK system (Keenan 1987). 
\footnote{For a detailed listing of spectral types and luminosity 
classes in the revised MK system see Keenan (1985).}
The majority of the stars were drawn from the following fundamental 
lists:
i) Morgan, Abt, and Tapscott (1978) 
for 29 stars O6--G0; ii) Keenan and McNeil (1989) for 45 stars G0--K5; 
and iii) Kirkpatrick, Henry, and McCarthy (1991) for 5 late--type dwarfs K5--M3.
We supplemented these primary standards with
an additional 5 secondary standards from the compilation of
Jaschek, Conde, and de Sierra (1964) and one late--type dwarf 
classified by Henry, Kirkpatrick, and Simons (1994). 

In order to cover as complete a range of stellar temperature
and luminosity as possible, we defined a two-dimensional grid 
with 26 bins of spectral type and three bins of luminosity class.  
Our temperature grid is 
binned $\times 2$ more coarsely in spectral subclass 
than the revised MK system, 
so that we typically sample only every other MK subclass.
The three luminosity bins are divided into
supergiants (I--II), giants (III), and subgiants/dwarf stars (IV--V).
A full sampling of this grid would have resulted in
78 distinct temperature/luminosity pairs. Our atlas 
includes a total of 85 sources with 53 of the bins filled. 
Grid coverage was finer among 
the later spectral types, where for stars GO and later, we filled
26 of the 27 bins (9 spectral types $\times$ 3 luminosity classes). 
In contrast, for stars earlier than GO, only 27 of the 51 bins were covered
(17 spectral types $\times$ 3 luminosity classes).

The 85 individual stars in our H--band survey are listed in Table~\ref{sample}
along with relevent stellar properties taken from the Bright Star Catalogue 
(Hoffleit \& Jaschek 1982). 
Additional restrictions on the sample selection included:
i) v$sini$ $<$ 250 km s$^{-1}$ with exception of HR3982 (B7 V); 
ii) near-solar metallicity (avoiding those MK standards which 
exhibit spectral peculiarities due to enhanced or deficient metal abundance); 
and iii) no visual companions within the beam (separations $1-5$'').
Our program was begun in advance of the K--band
FTS atlas of WH97, and their sources are drawn in large measure from
our sample. We note in Table~\ref{sample} the stars for which
K--band spectra can be found in the WH97 digital atlas.

\begin{figure}
\insertplot{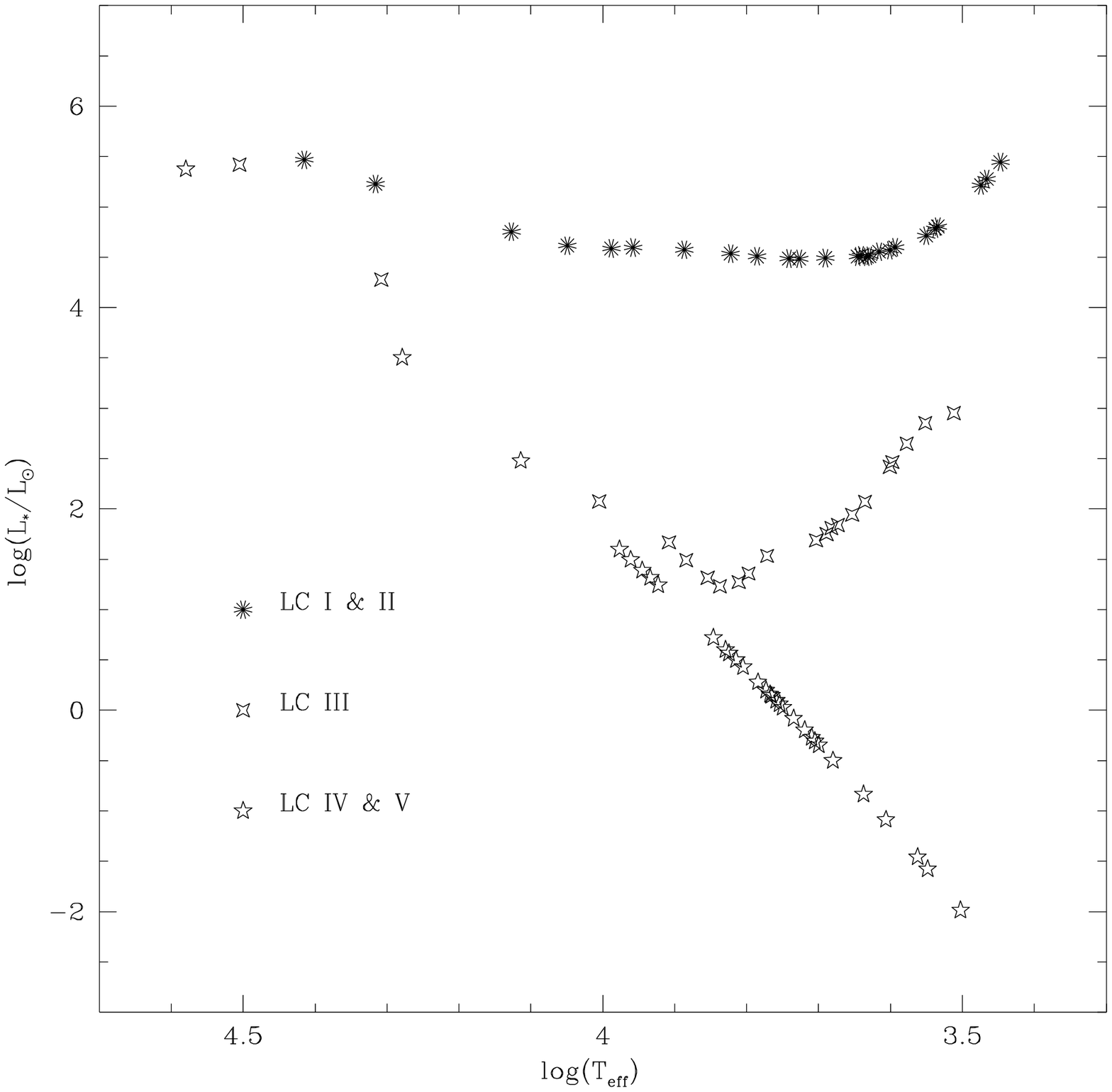}{8.0}{10.}{-0.75}{1.0}{0.8}{0}
\caption{Each star in our survey was transformed into the L$_*$ vs. T$_{eff}$
plane from the V magnitude listed in the Bright Star Catalog (Hoffleit \&
Jaschek 1982),
the spectral type--temperature calibration in Table~\ref{temp}, and bolometric
corrections for luminosity class I (applied to I--II), III, 
and V stars (applied to IV--V).}
\label{fig1}
\end{figure}

Table~\ref{temp} and Figure~\ref{fig1} provide additional insight into the
temperature and luminosity coverage of our sample. In Table~\ref{temp},
we list each of the spectral type and luminosity bins we have "filled".
For each bin in which there is at least one spectrum,  we give 
the corresponding effective temperature. For most stars, we adopted
the temperature scale of Tokunaga (1996), except for giants earlier than
G0 where we adopted Schmidt--Kaler (1982) 
\footnote{Recent work by Bessell, Castelli, and Plez (1998) provides 
updated temperatures, colors, and bolometric corrections for a wide
range of spectral types and luminosity classes.}. 
Figure~\ref{fig1} provides 
a schematic illustrating the temperature and luminosity coverage for
the 85 stars in our sample. In this illustration we have applied the 
same main sequence bolometric corrections to both dwarf (27) and subgiant 
(11) stars; as such they are indistinguishable in this diagram. 

\section{Observations and Data Calibration}

Observations of our 85 sample stars were obtained at the Mayall 4m telescope
at Kitt Peak National Observatory during four
separate observing runs from 1993--1994 (Table~\ref{ftslog}). We 
used the Fourier Transform Spectrometer (FTS) dual--output interferometer 
(Hall et al. 1979).   The FTS was ideal for this program
for several reasons.  First, the wavelength coverage of the 
FTS is limited only by the bandpass of the blocking filters, 
independent of the spectral resolution. This gave us complete coverage 
in the J-- and H--bands which would have been difficult to obtain with 
available grating spectrographs.  For example, our H--band spectral range 
is a factor of two greater than the spectra of 
Dallier et al. (1996).  Secondly, the spectral resolution is fixed by the path
difference scanned with the interferometer so we were able to 
chose the highest resolution possible and achieve S/N in excess of
75 for the majority of our sources.
 \footnote{For details concerning the advantages and disadvantages
of fourier transform spectroscopy, see Bell (1974).}  
Finally, because of the novel background 
subtraction algorithm of the 4m FTS described below, we were able to observe
the brightest stars in our sample ($H < 3.0^m$) during good daytime 
conditions (typically mornings). Combining daytime observations
with targeted nighttime observations of key faint 
sources, the FTS provided a uniform set of 
high quality spectra for a large sample of spectral standards.

Our observing program included simultaneous spectral coverage in both
the J--band and the H--band.  However, the
J--band data presented difficulties which
made it expeditious to focus our initial effort on the H--band. 
The primary problems with the J-band spectra were; i) the 
inherent difficulty in data reduction
due to rapid temporal variations
in telluric water vapor absorption; and ii) and the relative
paucity of strong features which would allow spectral
classification over the full range of stellar temperature and luminosity.
We defer discussion of the J--band spectra to a future contribution. 

Spectra were collected simultaneously in the J-- and H--bands
with the use of a dichroic beamsplitter to separate the wavelengths longward 
and shortward of 1.5 $\mu$m. 
%Four single element InSb detectors
%were used, two at each output for the J-- and H--bands respectively, 
%and cooled to 4K with liquid helium. 
Each star was centered 
within an input aperture of 3.8 arcsec while sky background was 
measured through an identical aperture 50.0 arcsec away.  
The interferogram was scanned at a rate of 1 kHz as the path 
difference was varied continuously from 0.0--0.75 cm providing an 
unapodized resolution of 0.8 cm$^{-1}$.  Data were obtained as
separate scan pairs, with the path difference varied first in one 
direction and then the other.  
%This produced two independent observations
%of each source.  
A forward--backward scan pair was treated as an 
``observation'' and observations were repeated in beam--switching 
mode (A--B--B--A).
% with the initial sky aperture becoming the source
%aperture in going from ``A'' to ``B'' mode and returning to the 
%original configuration in going from ``B'' to ``A''.
Because the sky background from each aperture produces 
an interferogram shifted in phase by 180$^{\circ}$ at each set of detectors, 
source spectra are background subtracted in fourier space
as they are collected.  This permits observations
of bright stars to be obtained during good daytime conditions.  These
beam--switched observations were repeated and scans were averaged 
until adequate signal--to--noise ratio (SNR) was achieved.  The interferograms 
were transformed at Kitt Peak National Observatory 
%using the software package 
%GRAMMY 
%\footnote{GRAMMY is based on a modified Cooley--Tukey algorithm for the 
%fast--fourier transform.} 
yielding spectra in units of relative flux versus wavenumber ($\sigma$ in cm$^{-1}$).
The transformed spectra were converted into fits format images and 
all data reduction was performed using the IRAF software package 
\footnote{IRAF is distributed by the National Optical Astronomy Observatories,
which is operated by the Association of Universities for Research in 
Astronomy, Inc., under contract to the National Science Foundation.}.
The spectra were then 
convolved with a gaussian filter of half--width $\delta = 1.2$ cm$^{-1}$.
This procedure, commonly referred to as apodization, 
eliminates ``ringing'' observed in the FTS spectra due to the finite 
scan path of the interferometer.  The resulting apodized resolution 
(Rayleigh criterion) was $\delta \sigma = 2.1$ cm$^{-1}$ 
giving a mean resolving power of $R=3000$ in the H--band.
At this stage the J-- and H--band spectra were separated for ease 
of reduction.  The slope of the continuum was normalized to 
1.0 using a four--segment spline fitting function. Care was taken to keep 
the residuals from this fit to within 1 \%.  

\begin{figure}
\insertplot{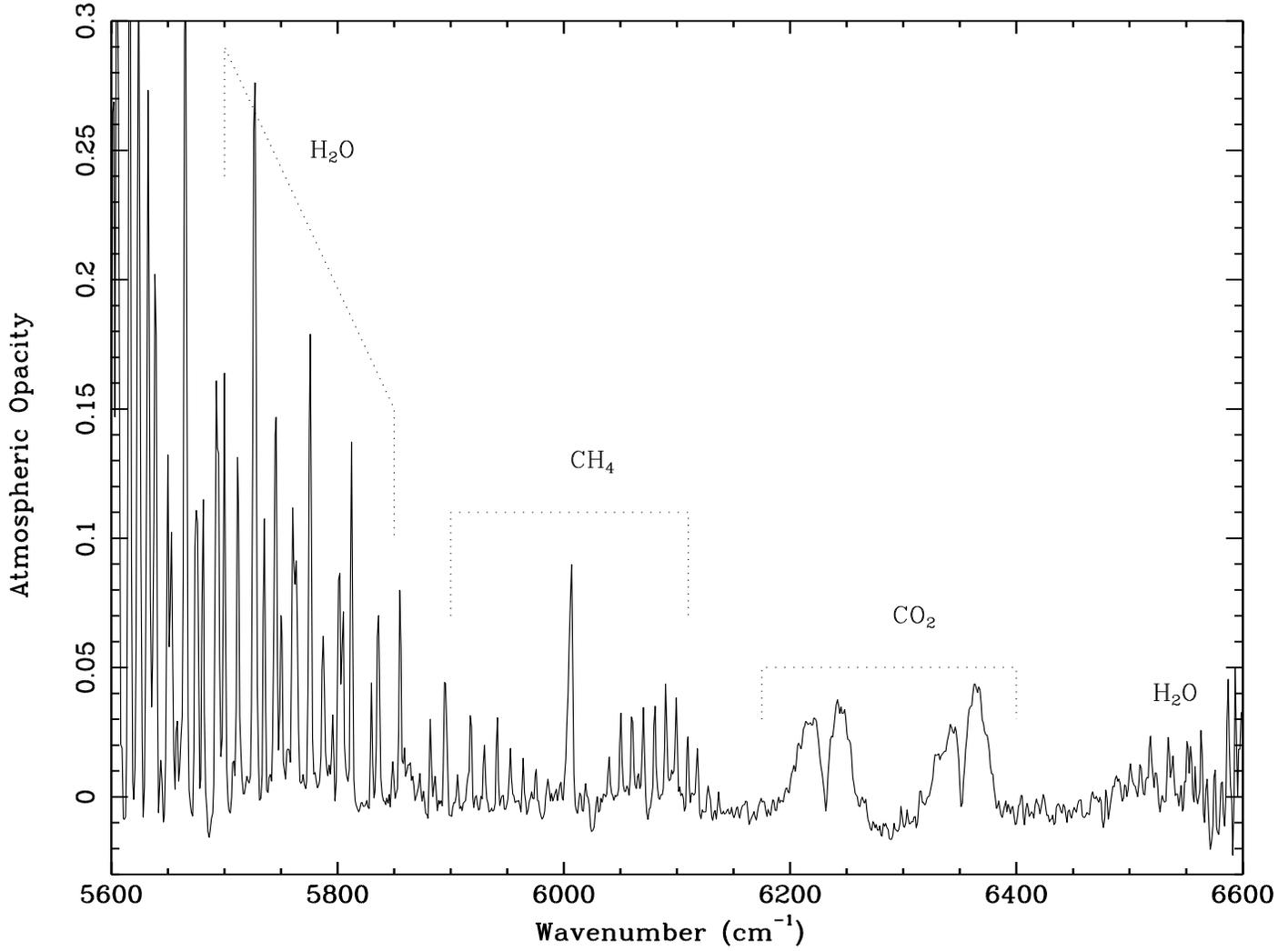}{8.0}{11.}{0.5}{0.00}{0.75}{1}
\caption{Atmospheric opacity in the H--band at a resolving power of $R=3000$.
The opacity was derived from ratios of high signal--to--noise spectra of the
same star observed at different airmass as discussed in the text.}
\label{fig2}
\end{figure}

Next we corrected the spectra for telluric absorption features present 
in the spectra which varied with zenith angle.  
We attempted to construct an opacity map for the earth's atmosphere by 
dividing normalized spectra obtained of the A0 star standards at different airmass.  
Because of the simplicity of the A0 star spectra, showing primarily hydrogen 
lines in absorption, it was relatively easy to monitor the degree to which 
this procedure was successful.  In dividing two normalized spectra of the same star 
taken at different airmass, 
all stellar photospheric absorption features should directly cancel, 
leaving only those absorption features due to the earth's atmosphere.
If we assume that the opacity of the telluric 
absorption is directly proportional to airmass we derive:
\begin{equation}
\tau(\sigma,X = 1.0) = \frac{1}{(X_{high}-X_{low})} \times ln[ \frac{I(\sigma,X_{low})}{I(\sigma,X_{high})}]
\end{equation}
where $\tau$ is the atmospheric opacity, $X_{low}$ is the low airmass value,
and $X_{high}$ is the high airmass value.  A typical opacity map derived in this 
way for the H--band is shown in Figure~\ref{fig2}.  
Several of the features in this map identified with known constituents of 
the earth's atmosphere such as water vapor, methane, and carbon dioxide, are
denoted in Figure~\ref{fig2}.  Again if the atmospheric opacity 
varies linearly with airmass we can simply scale the opacity for each 
star so that $\tau(\sigma,X) = X \times \tau(\sigma,X=1.0)$. Using this
technique we corrected the spectra to zero--airmass;
\begin{equation}
I(\sigma,X=0.0) = I(\sigma,X) \times e^{\tau(\sigma,X)}
\end{equation}
We used the highest signal--to--noise A0 standard star spectra 
($SNR > 100$) with the largest $\Delta X$ to define the opacity.  
We found some residual telluric absorptions, possibly due
to water vapor which do not vary strictly with airmass.
Such features severely complicate the reduction of the J--band spectra.

	Finally, the forward and backward scans of each star were
averaged and residuals of the differenced spectra were calculated
in order to evaluate the average SNR.  The observations were obtained with the goal 
of achieving SNR of 75 or greater.  In most cases this was achieved with 
the highest quality spectra reaching values of several hundreds. 
The average SNR for each stellar spectrum is included in 
Table~\ref{ftshewsup} --Table~\ref{ftshewdw} below.

\section{Line Identification and Dependence on Temperature and Luminosity}

\begin{figure}
\insertplot{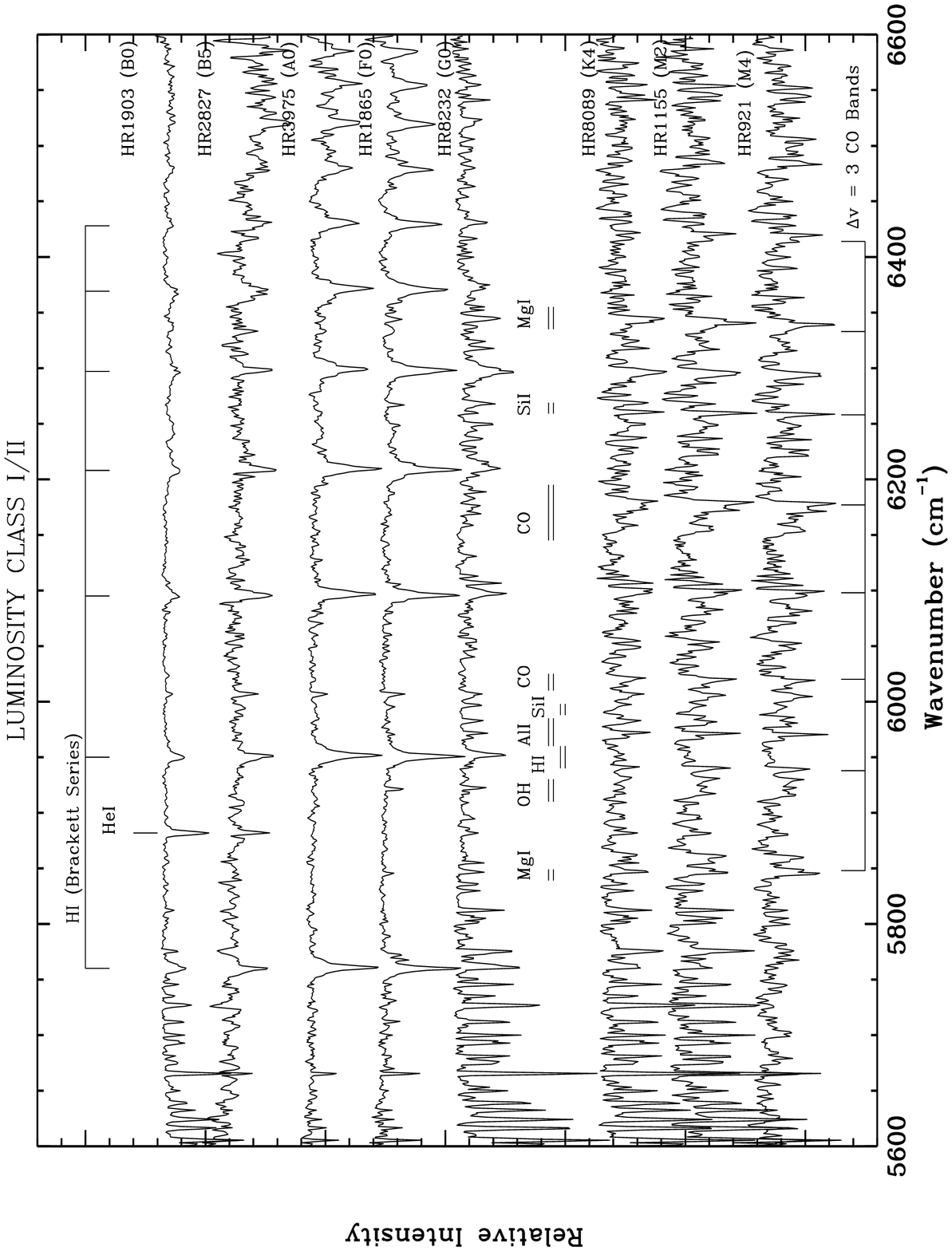}{8.0}{11.0}{-1.5}{1.0}{0.75}{0}
\caption{Representative H--band spectra of the MK standards plotted as a
function of effective temperature from high (top) to low (bottom) 
for luminosity class I--II stars.  Prominent features are indicated, including
the indices listed in Table~\ref{ftshband}.} 
\label{fig3}
\end{figure}

\begin{figure}
\insertplot{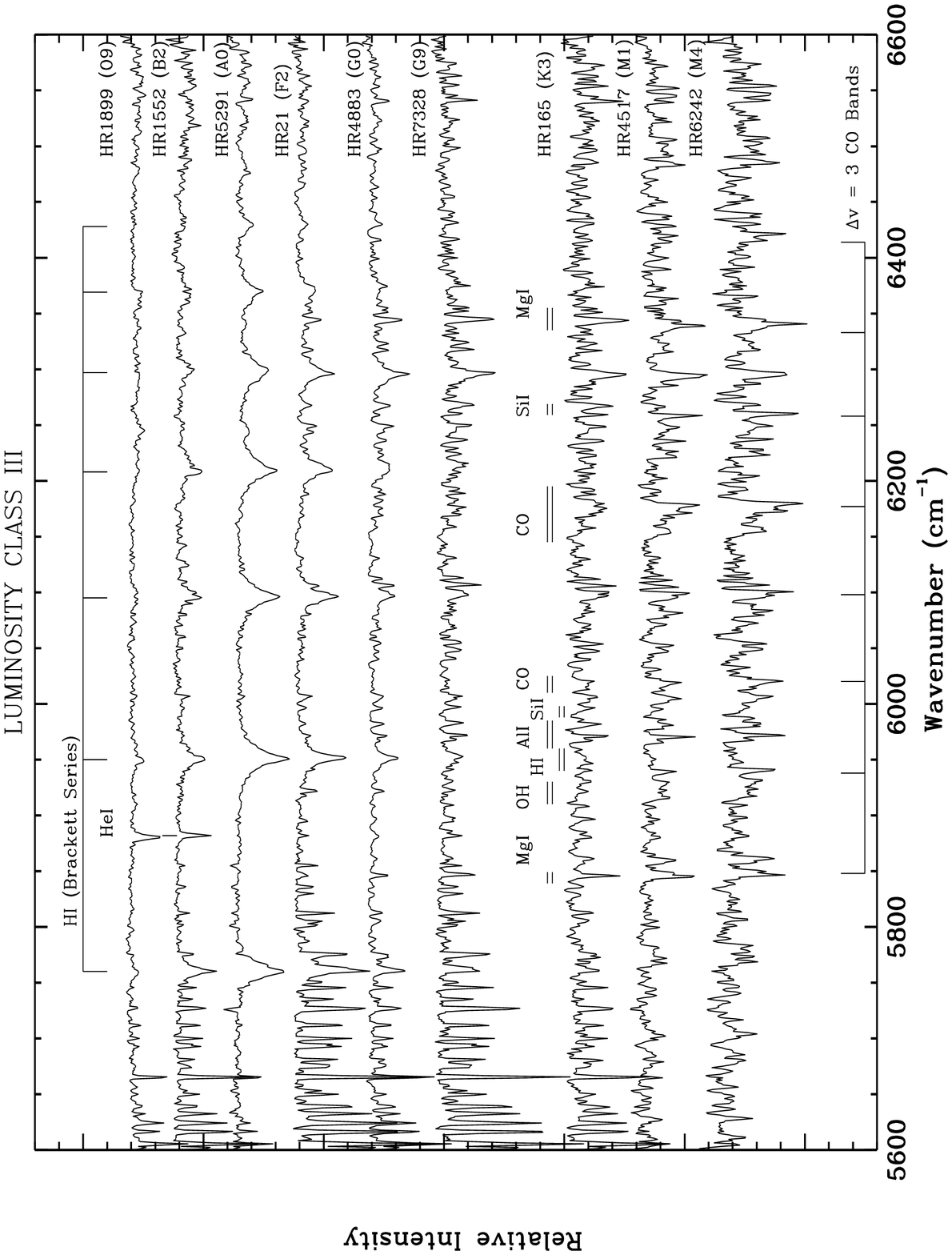}{8.0}{11.0}{-1.5}{1.0}{0.75}{0}
\caption{Representative H--band spectra of the MK standards plotted as a
function of effective temperature from high (top) to low (bottom) 
for luminosity class III stars.  Prominent features are indicated, including 
the indices listed in Table~\ref{ftshband}.} 
\label{fig4}
\end{figure}

\begin{figure}
\insertplot{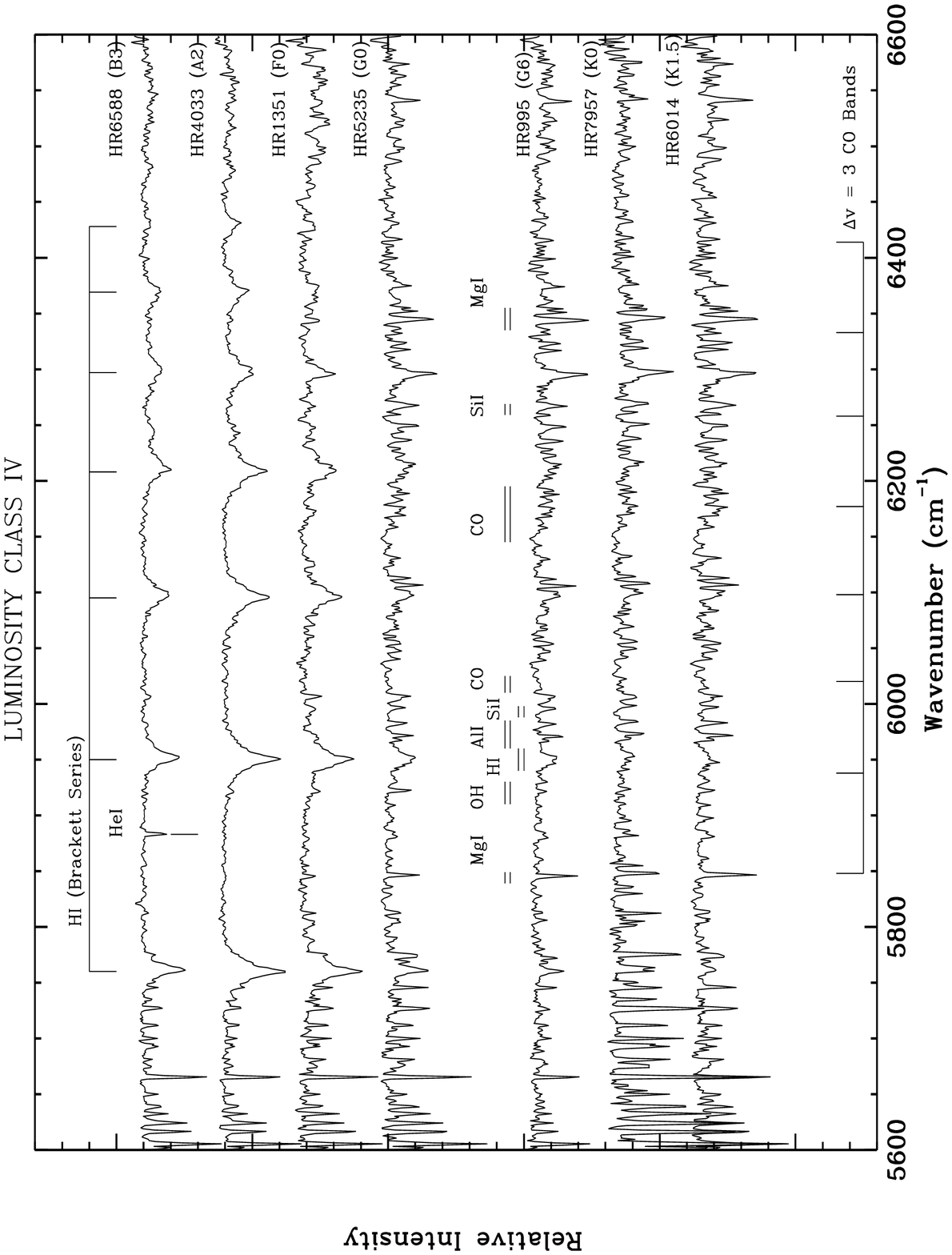}{8.0}{11.0}{-1.5}{1.0}{0.75}{0}
\caption{Representative H--band spectra of the MK standards plotted as a
function of effective temperature from high (top) to low 
(bottom) for luminosity class IV stars.  Prominent features are indicated, including 
the indices listed in Table~\ref{ftshband}.} 
\label{fig5}
\end{figure}

\begin{figure}
\insertplot{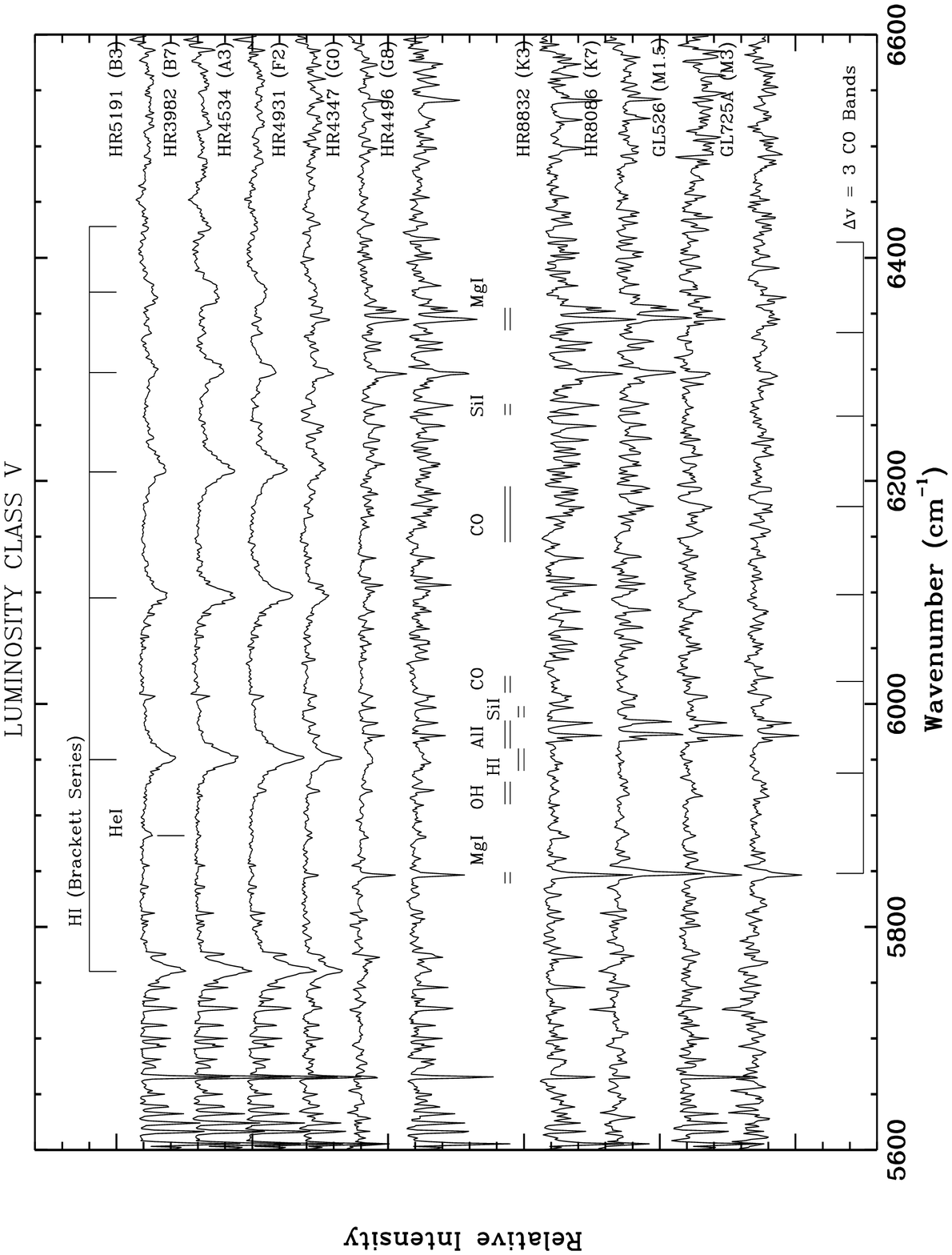}{8.0}{11.0}{-1.5}{1.0}{0.75}{0}
\caption{Representative H--band spectra of the MK standards plotted as a
function of effective temperature from high (top) to low (bottom) 
for luminosity class V stars.  Prominent features are indicated, including 
the indices listed in Table~\ref{ftshband}.} 
\label{fig6}
\end{figure}

Representative H--band spectra are shown in 
Figures~\ref{fig3}--~\ref{fig6} for luminosity classes
I--II, III, IV, and V, with prominent atomic and molecular features
identified. Line identifications were made for the strongest lines from
comparison with the solar photospheric and umbral near infrared atlases
(Livingston \& Wallace 1991 (LW91); Wallace \& Livingston 1992 (WL92)). 
However, at our moderate spectral resolution
many features are blended and we found 
the model atmosphere calculations of Oliva, 
Moorwood, and Origlia (1993) to be useful in identifying the 
dominant contributors to a blend in late--type stars. 

Visual inspection of the features in 
Figures~\ref{fig3}--~\ref{fig6}
reveals that $R=3000$ H--band spectra contain sufficent temperature
and luminosity sensitive features to enable spectral
classes to be distinguished.
Beginning with the early type stars, the dominant spectral features
are HeI 5882 cm$^{-1}$ (1.700 $\mu$m) and the
Brackett series of hydrogen from lines 4-10 (1.736 $\mu$m) to 4-16
(1.556 $\mu$m). The He I line exceeds the strength of the
Brackett lines in the very earliest stars (06 to B0), with a maximum
equivalent width of $\sim 0.83$ cm$^{-1}$ (HR1903; B0 Ia), and recedes to undetectable
levels ($\sim 0.10$ cm$^{-1}$) by spectral type B8.
From the late B to early F stars, the Brackett series dominates
the spectrum, after which lines of neutral atomic 
metals begin to take prominence.
The strongest metallic lines include MgI, SiI, CaI, AlI, and FeI,
which increase in strength toward the K stars. 
Finally molecular features of OH and CO dominate the 
spectra of the latest--type stars from K5--M5.  
The most striking luminosity-sensitive feature is the
second--overtone CO bandhead $[v, v^{'} = 6, 3]$ at 
6177 cm$^{-1}$ (1.619 $\mu$m), which is found in the 
spectra of the K and M stars. This feature is signficantly
stronger in stars of lower surface gravity at equivalent
spectral type.

To further enable spectral classification in the
H--band, we have identified a set of 9 features which are prominent
in stars of spectral type A-M. These include a relatively
isolated Brackett line (H4-11), 5 neutral metals, and 3 molecular
bands. In Table~\ref{ftshband}, we define 9 narrow 
band indices with bandpasses ranging from 10 to 50 cm$^{-1}$, which
include each of these features.  The variable widths of the bandpasses
were selected to minimize line blending,
contamination from residual telluric absorption,  
and sensitivity to radial velocity shifts.
Table~\ref{ftshband} also identifies the wavenumber of the dominant 
contributor and the lower state energy level, the
central wavenumber and passband of the index, and additional
species which may contribute to the index strength.
The equivalent widths of these 9 indices were
evaluated from the normalized spectra of our 85 survey stars, and
are tabulated in Tables ~\ref{ftshewsup} to ~\ref{ftshewdw} 
in units of cm$^{-1}$ 
\footnote{The conversion to angstroms 
is $EW(\AA) = [EW(cm^{-1})/ \sigma^2] \times 10^8$}.  
Uncertainties in these equivalent widths
depend on the SNR of the spectrum in question and the bandpass/strength
of the feature.  Errors range from 
$\sigma_{EW} = 0.02-0.1$ cm$^{-1}$ exceeding this upper limit in very few
cases. Multiple observations of several sources are listed
for comparison. 

\begin{figure}
\insertplot{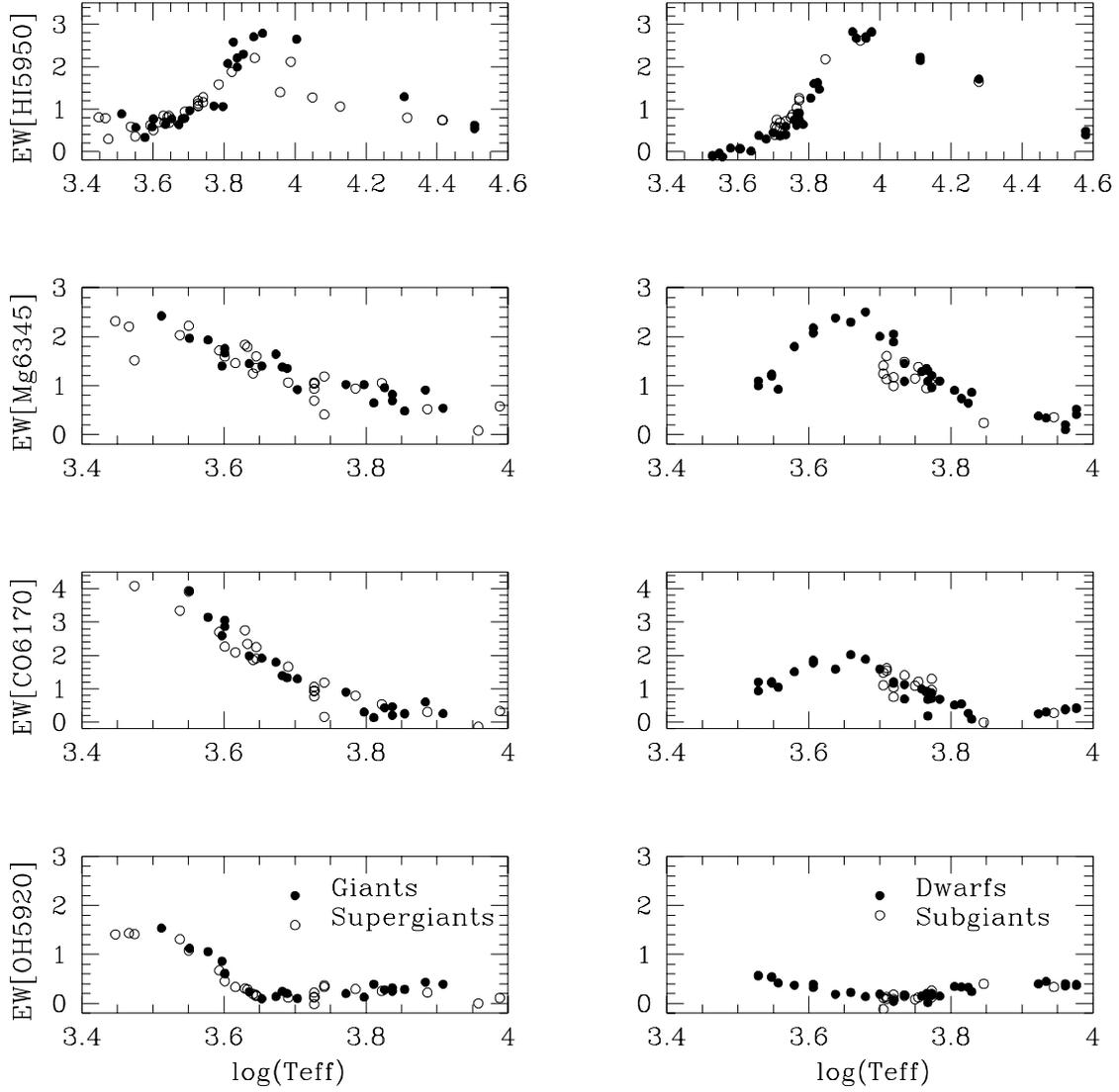}{8.0}{10.}{-0.75}{1.0}{0.8}{0}
\caption{Equivalent width of several species listed in Table~\ref{ftshband}
plotted as a function of effective temperature for stars of high 
(right panel) and low (left panel) surface gravity.
Typical errors are $< 0.1$ cm$^{-1}$.}
\label{fig7}
\end{figure}

The temperature and luminosity dependence for four representative
indices is illustrated in Figure~\ref{fig7}.
The 4--11 Brackett line (HI5950) behaves as expected,
with a rapid rise to a maximum (at a peak equivalent width 
of $\sim$ 3 cm$^{-1}$) as $T_{eff}$ approaches 10000 K, 
and a slower decline toward higher temperatures. The behavior of the
index is similar in both the dwarfs and the giants, although the
luminosity class I/II sources show a larger scatter, presumably due 
to intrinsic variability (e.g. Kaufer et al. 1996). 
The general behavior of the neutral atomic features is 
illustrated by the Mg6345 index. In luminosity classes IV--V,
this index reaches a maximum strength
between 5000-6000 K, with a peak equivalent width of
$\sim$ 2.5  cm$^{-1}$. In contrast, the maximum strength of this
index in the lower
surface gravity objects (also $\sim$ 2.5  cm$^{-1}$)
is found in the coolest stars in our sample,
monotonically decreasing toward higher temperatures; as
expected given the behavior of ionization state 
as a function of surface gravity. 
The two SiI indices exhibit similar behavior, but the AlI
index, (not shown) turns over at much lower temperatures in our dwarf
stars because of its lower ionization potential.  Note that 
we have chosen not to form an index based on the strongest SiI 
line at 6292 cm$^{-1}$ (1.5892 $\mu$m) because it is coincident 
with the 4--14 Brackett line of HI at 6297 cm$^{-1}$ (1.5881 $\mu$m).

The behavior of the molecular features is illustrated for both the
second--overtone $^{12}$CO (6,3) and the OH ($\Delta v = 2$)
indices. Both indices exhibit a similar behavior with
temperature and luminosity, becoming detectable
around  $T_{eff}$ = 5000, with a strength in the giants approximately
twice that in the dwarfs.  Similar behavior was noted by KH86 
in the first--overtone CO features in the K--band.   In 
dwarf stars the second--overtone CO index reaches a maximum before M5, and displays
a turnover toward the coolest stars. This may be due in part to 
features of CaI and FeI which contaminate the index for intermediate 
spectral types (F5--K3).  
Ali et al. (1995) find that the relationship between 
T$_{eff}$ and equivalent widths of the first--overtone CO bandheads 
flatten out between 3500--5000 K in dwarf stars.  From high resolution 
($R > 45,000$) FTS spectra, Wallace \& Hinkle (1996)
observed that the 2 $\mu$m continuum in late--type dwarf stars is suppressed by 
numerous water vapor features which are blended at intermediate to low resolution.
Predicting the equivalent widths of features where the apparent continuum is 
subject to temperature and luminosity effects is not straight--forward. 
In contrast, both the CO and OH indices
continue to rise at the coolest temperatures for stars of 
higher luminosity. However, the magnitude (and temperature) 
of the maximum in the dwarf stars differs between 
the CO and OH indices, which we use in the next section to define
a two-dimensional classification scheme for late--type stars. 
We note that the OH index begins to include a contribution from the
stark--broadened 4-11 Brackett line at 5949 cm$^{-1}$ creating a 
secondary maxiumum in the strength of this index around 10,000 K in
the dwarf stars.

While the temperature and luminosity dependence of the atomic features
is readily understood through application of 
the Saha and Boltzman equations governing the population of the
ionization states and energy levels respectively, 
the explanation behind the behavior of the
molecular features is more subtle. Two possibilities for the
factor of two enhancement in the molecular bands in the giants over
the dwarfs have been explored in the literature.
One attributes the luminosity dependence in
the molecular features to differing microturbulence in the atmospheres
of dwarfs and giants. The expectation is that larger microturbulence
in the lower surface gravity giants effectively
broadens the opacity of the feature over
a larger frequency interval in these saturated features, thereby 
enhancing the equivalent width (McWilliams \& Lambert 1984).
Another possible contributor is the differing
depth of the line formation region in the dwarfs versus the giants, 
which is fixed by the H$^{-}$ opacity.
As described by Gray (1992) higher surface gravity results in a higher
electron pressure (and thus H$^-$ column density).  This brings the CO line 
formation region closer to the stellar surface reducing $N_{CO}$
according to the following proportionality:
\begin{equation}
P_e \sim g^{1/3} \sim N_{H^-} \sim 1/N_{CO}
\end{equation}
In any case, this luminosity dependence of the band strength gives
an excellent empirical discriminant between giants and dwarfs, which
we exploit below to develop a two dimensional spectral index.

To discern surface gravity effects between the super--giants and giants or between 
sub--giants and dwarf stars requires more careful study. A detailed examination
of line strengths as a function of surface gravity at a fixed 
temperature reveal the expected trends. However, this behavior does not
reveal itself in the coarse analysis afforded by our narrow--band 
indices.  

\begin{figure}
\insertplot{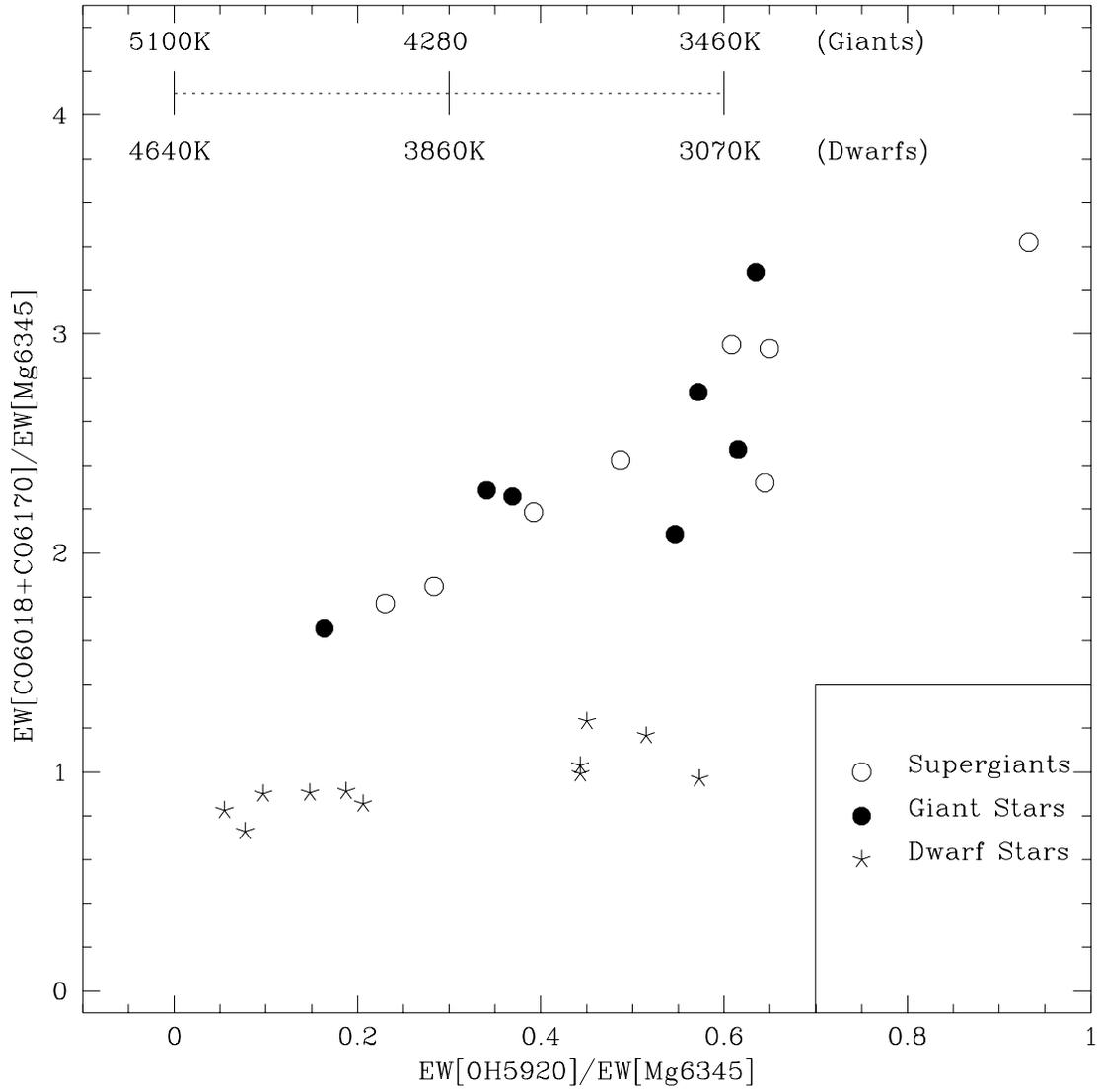}{8.0}{10.}{-0.75}{1.0}{0.8}{0} 
\caption{Two--dimensional spectral classification for late--type stars
K3--M5  using diagnostic line--ratios based on spectra with $SNR > 50$. 
The temperature scales derived for both dwarfs and giants are given.}
\label{fig8}
\end{figure}

While the temperature and luminosity sensitivities outlined above
can provide good spectral classification in many instances, 
discriminants which do not rely on absolute 
line strength are required when a star is
subject to near-infrared continuum veiling.
In this case line ratio diagnostics are to be preferred, since 
absorption features will appear shallower in the presence of
continuum excess but line ratios will be preserved as long as the
excess is not strongly wavelength dependent. 
We have identified one diagnostic based on line ratios 
which can be used to evaluate both temperature and luminosity 
for stars from K3-M5 in the presence of continuum veiling.
This two--dimensional spectral index is defined as:

\begin{equation}
\frac{EW[OH5920]}{EW[Mg6345]} \ vs. 
\ \frac{EW[CO6018+CO6170]}{EW[Mg6345]}
\end{equation}

\noindent
where EW is the equivalent width in cm$^{-1}$ for the indices
identified in Table~\ref{ftshband} and listed in 
Tables~\ref{ftshewsup}--~\ref{ftshewdw}.
The temperature and luminosity dependence of this diagnostic 
is illustrated in Figure~\ref{fig8}.  In this diagnostic, the 
ratio of the OH5920 to Mg6345 indices
is temperature sensitive, with distinct temperature
dependences for dwarfs and giants. Specifically we find

\begin{equation}
T_{eff} (V) = 4640 \pm 250 - (2610 \pm 110) \frac{EW[OH5920]}{EW[Mg6345]}
\end{equation}

\noindent
and

\begin{equation}
T_{eff} (III) = 5100 \pm 180 - (2730 \pm 80)\times \frac{EW[OH5920]}{EW[Mg6345]}
\end{equation}

\noindent
The comparison of this temperature sensitive ratio with the
sum of the two $^{12}$CO indices, also normalized to Mg6345,
then provides an excellent means of identifying both the temperature
and luminosity class of late--type stars.
Formal errors in the equivalent width suggest that 
spectral types can be evaluated to within $\pm 2$ subclasses
($\pm$ 300 K) from K3--M5 using spectra with $SNR > 50$ based on 
these indices alone.  

\section{Discussion and Summary}

Spectral classification in the near-infrared will become
increasingly important in the next decade, as the 
2MASS and DENIS near-infrared sky surveys reveal 
unprecented numbers of stars which are optically--invisible. 
Because the 1--2.5 $\mu$m region is on the Rayleigh--Jeans
tail of most stellar SEDs, it is not an ideal wavelength regime
to pursue spectral classification.  Yet, there are sufficient features
in both the H-- and the K--bands to allow most stellar photospheres
to be classified. For heavily reddened sources, the K--band will be the 
wavelength of choice. 
However, continuum emission from circumstellar dust with temperatures 
less than 2000 K
can heavily veil stellar photospheres at wavelengths 
greater than 2.0 $\mu$m.  In this case, shorter wavelength spectra 
are required in order to identify the underlying star.  

The early-type stars are probably the
most challenging for spectral classification in the
near--infrared. We find that a rough classification from 07-B8 
can be made in the H--band by the relative strengths of HeI 5882 cm$^{-1}$ 
and the Brackett series (see also Blum et al. 1997).
Hanson, Conti, and Rieke (1995) have established
a classification scheme in the K--band
for O--B stars relying on lines of helium as well as higher ionization species
obtained at $R > 1000$. 
For stars A through early K, the H--band may be superior to the
K band in providing a large number of intermediate 
ionization potential species with strong features such as MgI, SiI, and FeI
in addition to the numerous Brackett series features. 
Stars K3--M5 are probably best classified in the K--band 
(KH86; Ali et al. 1995; WH97) using atomic features of MgI, CaI, and NaI 
as well as the first--overtone CO bandheads observed at $R \sim 1000$. 
However we have found that
these stars also have strong temperature and luminosity sensitive
features in the H--band such as MgI, AlI, OH, and the 
second--overtone CO bandheads.  
The very latest--type stars ($> M5$) have very strong, broad, molecular 
features which can be identified at resolutions as low as $R \sim 300$. 
Kirkpatrick et al. 1993 (see also Jones et al. 1996) have classified stars 
in the I-- or the J--band employing 
features due to VO, TiO, and FeH.  In addition, 
broad water vapor bands observed throughout the 1--2.5 $\mu$m region 
(Jones et al. 1995) are an important opacity source in the atmospheres 
of the coolest stars as well as brown dwarfs (Allard \& Hauschildt 1995).
While the I-- or J--bands
are probably the best spectral regions to classify extremely cool stars
(as they lie on the Wien side of the Planck function for these objects) more 
heavily obscured objects can still be profitably observed at low resolution 
in the J-- and H--bands (e.g. NICMOS on HST) or in the K--band 
(Wilking, Greene, \& Meyer 1998) in search of these water vapor absorptions. 

The H--band spectral atlas we have presented is comprised of moderate
resolution spectra with $R \sim 3000$. In contrast, most 
spectral classification is typically carried out with $R \sim 500-1000$.
The strongest and broadest features in the H--band are the CO(6-3) bands
and the Brackett lines. These features could be identified with 
much lower spectral resolution than our survey, at $R \sim 500$.
The most crowded region in the H--band spectra
is that in the vicinity of the HI line at 5948.50 cm$^{-1}$ (1.68110 $\mu$m), 
the AlI triplet at 5964--5980 cm$^{-1}$ (1.677--1.672 $\mu$m), 
and the SiI line at 5993.29 cm$^{-1}$ (1.66853 $\mu$m). 
In order to properly separate these important features from each 
other, a resolving power of $R \sim 1000$ is required.  At this 
resolution, one can also obtain measurements of the HeI 
line at 5882 cm$^{-1}$ (1.700 $\mu$m). 
At $R = 3000$ one can resolve individual components of the AlI 
triplet, the Mg I doublet, and the 
CO bandheads, as well as the stark--broadened Brackett lines in the 
early--type dwarf stars (Table~\ref{ftshband}).
An additional issue in the near--infrared is the significant contribution
to shot-noise from air-glow lines. In the H--band air--glow from OH
is sufficently bright and variable that they
compromise $R=1000$ H--band spectral classification
for very faint sources. Spectral resolution as high as
$R \sim 5000$ will be required to resolve the bulk of these air--glow features
and to obtain adequate SNR spectra of faint objects 
\footnote{ See Herbst (1994) for a comprehensive discussion of OH 
airglow background supression strategies.}.

In summary, we present an H--band spectral atlas at a resolving 
power of $R = 3000$ that
spans a wide range in stellar temperature (O7-M5) and 
luminosity class (I-V). This spectral region contains a number of 
temperature and/or luminosity sensitive
atomic and molecular features which will allow spectral classification
to be carried out in the H-band.  As an example of the efficacy of this 
spectral range for distinguishing stellar spectral types, we define a set of 
narrow--band indices which, with $ SNR \sim 50$, permit
classification of late--type stars on the MK system within $\pm 2$ subclasses.
It appears however, that for most applications obtaining H--band spectra at
$R \sim 1000$ will be sufficient for classification. 

\section{Appendix A:  Electronic Availability of the Data}

The final reduced averaged spectra as well as the difference of the 
forward and 
backward scan pairs (see Section 3 for description of the reduction 
procedure) are available through the Astronomical Data Center (ADC) 
for each observation listed in this paper. The ADC can be 
contacted directly: i) by post at  Astronomical Data Center, NASA Goddard 
Space Flight Center, Code 631, Greenbelt, MD  20771; or 
ii) by telephone at (301) 286--8310; or iii) by fax at (301) 286--1771; 
or iv) via the internet at  http://adc.gsfc.nasa.gov.  The data
are in fits format with pertinent header information included for each 
image.  These fits format files, useful plotting 
routines, and other relevent information are also available on the World Wide 
Web at http://donald.phast.umass.edu.
The raw FTS data are also available directly from NOAO 
(contact KHH for details). 

\acknowledgments
 
We would like to thank Lori Allen, Ed Chang, Lynne Hillenbrand,
Susan Kleinmann, 
Michael Skrutskie, and Lloyd Wallace for helpful discussions.
Special thanks to John Carpenter for assisting in the initial compilation
of the standard star lists, and to Karen Strom and Stephen Friedman 
for their assistance in
making the data available electronically. Antonella Romano  
provided assistance in preparing the tables and figures
for publication.  Support for MRM during the final stages of this
work was provided by NASA through Hubble Fellowship grant 
\# HF--01098.01--97A awarded by the Space Telescope Science 
Institute, which is operated by the Association of Universities
for Research in Astronomy, Inc., for NASA under contract NAS 5--26555. 
SE acknowledges support from the National Science Foundation's 
Faculty Award for Women Program.  This work was supported in part
through a grant from the National Science Foundation (\# AST--9114863) 
to SES.

\vfill\eject

%%%%%%%%%%%%%%%%%%%%%%%%%%%%%%%%%%%%%%%%%%%%%%%%%%%%%

\vfill\eject

\vfill\eject

%%%%%%%%%%%%%%%%%%%%%%%%%%%%%%%%%%%%%%%%%%%%%%%%%%%%%

\begin{tiny}

\begin{table}
\caption{H--Band Survey Sample}
\begin{tabular}{lllllll}
\tableline
\multicolumn{1}{c}{HR \#}    &
\multicolumn{1}{c}{NAME \tablenotemark{a}} & 
\multicolumn{1}{c}{$m_H$ \tablenotemark{b}} & 
\multicolumn{1}{c}{ST}   &
\multicolumn{1}{c}{Class}  &
\multicolumn{1}{c}{RV (km s$^{-1}$) \tablenotemark{c}} & 
\multicolumn{1}{c}{vsini (km s$^{-1}$) }  \\\tableline\tableline
1903     & *46 $\epsilon$ Ori & 2.4    & B0         & Ia         & 26SB     & 87        \\  
1203     & 44 $\zeta$ Per    & 3.4    & B1         & Ib         & 20SB     & 59        \\  
2827     & 31 $\eta$ CMa     & 2.5    & B5         & Ia         & 41V      & 45        \\  
1713     & *19 $\beta$ Ori    & 0.1    & B8         & Ia         & 21SB     & 33        \\  
3975     & *30 $\eta$ Leo     & 3.3    & A0         & Ib         & 3V       & 20        \\  
7924     & *50 $\alpha$ Cyg   & 1.0    & A2         & Ia         & -5SBO    & 21        \\  
1865     & 11 $\alpha$ Lep   & 2.0    & F0         & Ib         & 24       & 13        \\  
1017     & *33 $\alpha$ Per   & 0.9    & F5         & Ib         & -2V      & 18        \\ 
7796     & *37 $\gamma$ Cyg   & 1.1    & F8         & Ib         & -8V      & 20        \\ 
8232     & *22 $\beta$ Aqr    & 1.5    & G0         & Ib         & 7V?      & 18        \\ 
7479     & 5 $\alpha$ Sge    & 2.4    & G1         & II         & 2V?      & 0         \\ 
7063     & $\beta$ Sct       & 2.2    & G4         & IIa        & -22SB10  & 10        \\ 
8752     & *-                & 3.6   &  G4v        & $>I$ \tablenotemark{d} & -58V?    & 35 \\ 
7314     & *21 $\theta$ Lyr   & 2.1    & K0         & +II        & -31V     & $<$19     \\ 
6713     & 93 Her            & 2.4    & K0.5       & IIb        & -24      & $<$17     \\ 
8465     & *21 $\zeta$ Cep    & 1.1    & K1.5       & Ib         & -18SB    & $<$17     \\ 
6498     & 49 $\sigma$ Oph   & 1.9    & K2         & II         & -27      & $<$19     \\  
603      & 57 $\gamma^1$ And & -0.5   & K3         & -IIb       & -12SB    & $<$17     \\ 
8089     & *63 Cyg            & 1.5    & K4         & Ib-IIa     & -26V     & --        \\    
8079     & *62 $\xi$ Cyg      & 0.5    & K4.5       & Ib-II      & -20SB    & $<$17     \\ 
2061     & 58 $\alpha$ Ori   & -2.?   & M1-2       & Ia-Iab     & 21SB     & --        \\ 
1155  & *--                & 0.5    & M2         & +IIab      & -3V      & --        \\ 
921      & 25 $\rho$ Per     & -1.7   & M4         & II         & 28       & --        \\ 
7009  & *--                & 0.6    & M4.5-M5    & +II        & -19      & --        \\ 
6406  & *64 $\alpha^1$ Her & -2.4   & M5         & Ib-II      & -33V     & 21        \\\tableline
1899     & 44 $\iota$ Ori    & 3.5    & O9         & III        & 22SB2O   & 130       \\ 
1552     & 3 $\pi^4$ Ori     & 4.1    & B2+B2      & III        & 23SBO    &  40       \\ 
5291     & *11 $\alpha$ Dra   & 3.5    & A0         & III        & -13SBO   & 18        \\
403      & *37 $\delta$ Cas   & 2.3    & A5         & III-IV     & 7SB      & 113       \\ 
1412     & *78 $\theta^2$ Tau & 2.9    & A7         & III        & 40SB1O   & 78        \\ 
4031     & *36 $\zeta$ Leo    & 2.8    & F0         & III        & -16SB    & 84        \\ 
21       & * 11 $\beta$ Cas   & 1.6    & F2         & III--IV    & 12SB     & 70        \\ 
5017     & *20 CVn            & 3.9    & F3         & III        & 8V?      & 17        \\ 
2706     & 48 Gem            & 5.0    & F5         & III-IV     & 13V      & 74        \\ 
8905     & *68 $\upsilon$ Peg & 3.3    & F8         & III        & -11      & 79        \\ 
4883     & *31 Com            & 3.0    & G0         & III        & -1V?     & 77        \\ 
4716     & 5 CVn             & 2.8    & G6         & III        & -12SB    & $<$17     \\ 
7328     & 1 $\kappa$ Cyg    & 1.6    & G9         & III        & -29SB    & $<$17     \\ 
7949     & 53 $\epsilon$ Cyg & 0.2    & K0         & -III       & -11SB?   & $<$17     \\ 
8317     & *11 Cep            & 2.3    & K1         & III        & -37      & $<$17     \\ 
6299     & *27 $\kappa$ Oph   & 0.8    & K2         & III        & -56V     & $<$17     \\ 
165      & 31 $\delta$ And   & 0.5    & K3         & III        & -7SB1O   & $<$17     \\ 
6705     & *33 $\gamma$ Dra   & -1.2   & K5         & III        & -28      & $<$17     \\ 
152      & --                & 1.7    & K5-M0      & III        & -33V     & $<$17     \\ 
4517     & *3 $\nu$ Vir       & 0.3    & M1         & IIIab      & 51V?     & --        \\ 
6242     & --                & 0.4    & M4         & +III-IIIa  & -7V?     & --        \\ 
7886     & *--                & -0.6   & M6         & III        & -66V?    & --        \\\tableline  
\end{tabular}
\label{sample} 
\tablenotetext{a}
{ ``*'' indicates that the object also appears in the K--band atlas of WH97.}
\tablenotetext{b}{H--band magnitudes estimated from V--band magnitude from the Bright Star
Catalogue, spectral type, and intrinsic colors (Koornneef, 1983).}
\tablenotetext{c}{ ``V'' and ``V?'' indicate variable or suspected variable radial velocity respectively.
``SB1'' and ``SB2'' indicate single-- and double--lined spectroscopic binaries respectively.
``O'' indicates that orbital data is available in the Bright Star Catalogue.}
\end{table}

\vfill\eject

\begin{table}
\begin{tabular}{lllllll}
\tableline
\multicolumn{1}{c}{HR \#}    &
\multicolumn{1}{c}{NAME \tablenotemark{a}}   &
\multicolumn{1}{c}{$m_H$ \tablenotemark{b}} &
\multicolumn{1}{c}{ST}   &
\multicolumn{1}{c}{Class}  &
\multicolumn{1}{c}{RV (km s$^{1}$) \tablenotemark{c}}     &
\multicolumn{1}{c}{vsini (km s$^{1}$) }  \\\tableline\tableline
6588     & 85 $\iota$ Her    & 4.3    & B3         & IV         & -20SB1O  & 11        \\ 
4033     & *33 $\lambda$ UMa  & 3.3    & A2         & IV         & 18V      & 48        \\ 
1351     & 57 Tau            & 4.9    & F0         & IV         & 42SB1?   & 109       \\ 
5235     & 8 $\eta$ Boo      & 1.5    & G0         & IV         & 0SB1O    & 13        \\ 
5409     & 105 $\phi$ Vir    & 2.8    & G2         & IV         & -10SB    & 0         \\ 
6623     & 86 $\mu$ Her      & 1.4    & G5         & IV         & -16V     & 20        \\ 
995      & 59 Ari            & 3.9    & G6         & IV         & 0V       & --        \\ 
7602     & 60 $\beta$ Aql    & 1.7    & G8         & IV         & -40V     & $<$ 16    \\ 
7957     & *3 $\eta$ Cep      & 1.2    & K0         & IV         & -87      & $<$17     \\ 
5901     & 11 $\kappa$ CrB   & 2.5    & K1         & IVa        & -24      & $<$17     \\ 
6014     & --                & 3.6    & K1.5       & IV         & -4V      & --        \\\tableline
2456     & *15 Mon            & 5.5    & O7         & V(e)       & 33SB     & 63        \\ 
5191     & *85 $\eta$ UMa \tablenotemark{e} & 2.4    & B3         & V          & -11SB?   & 205       \\ 
3982     & *32 $\alpha$ Leo \tablenotemark{e}   & 1.6    & B7         & V          & 6SB      & 329       \\ 
7001     & * $\alpha$ Lyr     & 0.0    & A0         & V          & -14V     & 15        \\ 
2491     & *9 $\alpha$ CMa    & -1.5   & A1         & Vm         & -8SBO    & 13        \\ 
4534     & *94 $\beta$ Leo \tablenotemark{e}   & 2.0    & A3         & V          & 0V       & 121       \\ 
4357     & *68 $\delta$ Leo \tablenotemark{e}  & 2.3    & A4         & V          & -20V     & 181       \\ 
4931     & 78 UMa            & 4.1    & F2         & V          & -10V?    & 92        \\ 
1279     & --                & 5.1    & F3         & V          & 36SB1?   & 25        \\ 
2943     & *10 $\alpha$ CMi   & -0.6   & F5         & IV-V       & -3SBO    & 6         \\ 
1538     & 59 Eri \tablenotemark{e}           & 2.3    & F6         & V          & 35       & --        \\ 
4375     & *53 $\xi$ UMa      & 3.0    & G0         & V          & -16SB1O  & 1         \\ 
4983     & 43 $\beta$ Com    & 3.1    & F9.5       & V          & 6SB?     & 6         \\ 
483      & *--                & 3.7    & G1.5       & V          & 4V?      & 2         \\ 
4374     & 53 $\xi$ UMa      & 3.5    & G0         & V          & -16SB1O  & 3         \\ 
5072     & 70 Vir            & 3.6    & G4         & V          & 5V       & 1         \\ 
4496     & *61 UMa            & 3.8    & G8         & V          & -5V      & $<$17     \\ 
7462     & *61 $\sigma$ Dra   & 3.0    & K0         & V          & 27V      & $<$17     \\ 
1084     & *18 $\epsilon$ Eri & 1.6    & K2         & V          & 15V?     & $<$17     \\ 
8832     & *--                & 3.2    & K3         & V          & -18V     & --        \\ 
--       & GL570A            & 3.0    & K4         & V          & -        & -         \\ 
8085     & *61 Cyg            & 2.4    & K5         & V          & -64V     & $<$17     \\ 
8086     & 61 Cyg            & 3.1    & K7         & V          & -64V?    & =$<$25    \\ 
--       & GL338A            & 4.5    & M0         & V          & -        & -         \\ 
--       & GL526 \tablenotemark{f} & 4.5    & M1.5    & V          & -        & -         \\ 
--       & *GL411             & 3.6    & M2         & V          & -        & -         \\ 
--       & *GL725A            & 4.7    & M3         & V          & -        & -         \\\tableline
\end{tabular}
\tablenotetext{d}{Previously classified as G0Ia, but it is intrinsically brighter.} 
\tablenotetext{e}{Taken from Jaschek, Jaschek, and Condi (1964).}
\tablenotetext{f}{Taken from Henry, Kirkpatrick, and Simons (1994).}
\end{table}

\clearpage 
\vfill\eject

\begin{table}
\caption{Temperature Bins for Survey Stars}
\begin{tabular}{rlll}
\tableline
\multicolumn{1}{c}{Spectral Type \tablenotemark{a}} &
\multicolumn{1}{c}{T$_{I-II}$ \tablenotemark{b}}    &
\multicolumn{1}{c}{T$_{III}$ \tablenotemark{b} \tablenotemark{c}}     &
\multicolumn{1}{c}{T$_{IV-V}$ \tablenotemark{b}}    \\\tableline
O6-O8--	 & \nodata  & 37000   &  38000   \\
O9--	 & 32500    & 32000   &  33200   \\
O9.5--	 & \nodata  & \nodata &  31450   \\
B0--	 & 26000    & 29000   &  29700   \\
B1--	 & 20700    & 24000   &  25600   \\
B2--	 & 17800    & 20300   &  22300   \\
B3--	 & 15600    & 17100   &  19000   \\
B4	 & 13900    & \nodata &  17200   \\
B5--	 & 13400    & 15000   &  15400   \\
B6	 & 12700    & 14100   &  14100   \\
B7--	 & 12000    & 13200   &  13000   \\
B8--	 & 11200    & 12400   &  11800   \\
B9--	 & 10500    & 11000   &  10700   \\
A0--	 &  9730    & 10100   &   9480   \\
A1--	 &  9230    &  9480   &  \nodata \\
A2	 &  9080    &  9000   &   8810   \\
A5--	 &  8510    &  8100   &   8160   \\
A7--	 & \nodata  &  7650   &   7930   \\
F0--	 &  7700    &  7150   &   7020   \\
F2--	 &  7170    &  6870   &   6750   \\
F5--	 &  6640    &  6470   &   6530   \\
F7--	 & \nodata  & \nodata &   6240   \\
F8--	 &  6100    &  6150   &  \nodata \\
G0--	 &  5510    &  5910   &   5930   \\
G2 	 & \nodata  & \nodata &   5830   \\ 
G3	 &  4980    & \nodata &  \nodata \\
G4	 & \nodata  &  5190   &   5740   \\
G6	 & \nodata  &  5050   &   5620   \\
G8--	 &  4590    &  4960   &  \nodata \\
K0--	 &  4420    &  4810   &   5240   \\
K1--	 &  4330    &  4610   &  \nodata \\
K2--	 &  4260    &  4500   &   5010   \\
K3--	 &  4130 \tablenotemark{d} &  4320   &  \nodata \\
K4--  	 & \nodata  &  4080   &   4560   \\
K5--	 &  3850 \tablenotemark{d} &  3980   &   4340   \\ 
K7	 & \nodata  & \nodata &   4040   \\
M0--	 &  3650 \tablenotemark{d} &  3820   &   3800   \\
M1--	 &  3550 \tablenotemark{d} &  3780   &   3680   \\
M2--	 &  3450 \tablenotemark{d} &  3710   &   3530   \\
M3--	 &  3200 \tablenotemark{d} &  3630   &   3380   \\
M4--	 &  2980 \tablenotemark{d} &  3560   &   3180   \\
M5--	 & \nodata  &  3420   &   3030   \\
M6--	 & \nodata  &  3250   &   2850   \\\tableline 
\end{tabular}
\label{temp}
\tablenotetext{a}{``--'' denotes full sub--class in the revised MK system.  
Other full sub--classes for which temperatures are not available without 
interpolation include B0.5, A3, A8, F3, F9, and G5.} 
\tablenotetext{b}{Data taken from Tokunaga, 1996, Astrophysical Quantities, in press.}
\tablenotetext{c}{For giant stars G0 and later data taken from Schmidt--Kaler (1982).} 
\tablenotetext{d}{Temperature for star of Luminosity Class Iab.} 
\end{table}

\end{tiny}

\clearpage 
\vfill\eject

\begin{small}
\begin{table}
\caption{Journal of Observations for the H--Band Survey}
\begin{tabular}{lll}
\tableline
\multicolumn{1}{c}{Date} &
\multicolumn{1}{c}{Type} &
\multicolumn{1}{c}{ \# of Stars} \\\tableline\tableline
March 9--10, 1993 & Day & 17 \\
April 1--3, 1993 & Day/Night & 30 \\
May 18--19, 1993 & Day & 10 \\
January 30--31, 1994 & Day/Night & 42 \\\tableline
\end{tabular}
\label{ftslog}
\end{table}

\clearpage 
\vfill\eject

\begin{table}
\caption{H--Band Spectral Indices for Classification of A-M Stars} 
\begin{tabular}{lllllll}
\tableline
\multicolumn{1}{c}{Main Feature \tablenotemark{a}}&
\multicolumn{1}{c}{E$_{low}$ (ev) } &
\multicolumn{1}{c}{$\sigma$ (cm$^{-1}$) \tablenotemark{a}}&
\multicolumn{1}{c}{$\lambda$ ($\mu$m) } &
\multicolumn{1}{c}{$\sigma_{cent}$ (cm$^{-1}$)}&
\multicolumn{1}{c}{$\Delta \sigma$ (cm$^{-1}$)} &
\multicolumn{1}{c}{Other contr. \tablenotemark{a}} \\\tableline\tableline
MgI(4s--4p)          & 5.39 \tablenotemark{b}& 5843.41 & 1.71133 & 5844   & 10 & CO, Fe, Ni, OH  \\
OH($\Delta v = 2$) & 0.76 \tablenotemark{c}& 5920:   & 1.689   & 5920   & 20 & C, CO, Fe, Ni   \\
HI(4-11)             & 12.75 \tablenotemark{d}& 5948.50 & 1.68110 & 5950   & 20 & CO, Fe, Ni, Si \\
AlI(4p--4d tr)       & 4.09 \tablenotemark{e}& 5963.76 & 1.67679 & 5972.5 & 25 & CO, Fe, Ni, OH    \\
                     &      & 5968.31 & 1.67552 &        &    &     \\
                     &      & 5979.60 & 1.67235 &        &    &     \\
SiI(4p--3d)          & 5.98 \tablenotemark{f}& 5993.29 & 1.66853 & 5993   & 10 & CO, Fe, Ni, OH   \\
$^{12}$CO(8,5)bh     & 1.55 \tablenotemark{g}& 6018    & 1.662  & 6017.5 & 15 & Fe, OH, S   \\
$^{12}$CO(6,3)bh     & 1.05 \tablenotemark{g} & 6177    & 1.619  & 6170   & 50 & Ca, Fe, Ni, OH, Si      \\
SiI(4p--5s)          & 5.98 \tablenotemark{f}& 6263.92 & 1.59644 & 6264   & 10 & Fe, Mg, Ni, OH       \\
MgI(4s--4p tr)       & 5.93 \tablenotemark{b}& 6341.10 & 1.57701 & 6345   & 20 & CN, CO, Fe, H$_2$O, Ni, OH \\
                     &      & 6347.88 & 1.57533 &        &    &      \\
                     &      & 6351.22 & 1.57450 &        &    &      \\\tableline
\end{tabular}
\tablenotetext{a}{Species indentification and frequencies from LW91 and WL92.}
\tablenotetext{b}{Lower state energy level from Risberg (1965).} 
\tablenotetext{c}{Coxon and Forester (1982).} 
\tablenotetext{d}{Garcia and Mack (1965).} 
\tablenotetext{e}{Eriksson and Isberg (1963).} 
\tablenotetext{f}{Litzen (1964).} 
\tablenotetext{g}{George, Urban, \& LeFloch (1994).} 
\label{ftshband}
\end{table}
\end{small}

\clearpage 
\vfill\eject

\begin{tiny}

\begin{table}
\caption{Equivalent Widths of Spectral Indices for Luminosity Class I--II Stars}
\begin{tabular}{llllllllllll} 
\tableline 
\multicolumn{1}{c}{Source} & 
\multicolumn{1}{c}{T$_{e}$ } & 
\multicolumn{1}{c}{SNR} & 
\multicolumn{1}{c}{Mg5844} & 
\multicolumn{1}{c}{OH5920} & 
\multicolumn{1}{c}{HI5950} & 
\multicolumn{1}{c}{Al5973} & 
\multicolumn{1}{c}{Si5993} & 
\multicolumn{1}{c}{CO6018} & 
\multicolumn{1}{c}{CO6170} & 
\multicolumn{1}{c}{Si6264} & 
\multicolumn{1}{c}{Mg6345} \\\tableline\tableline 
HR1903   & 26000 & 203 & -0.03 &  0.09 &  0.75 &  0.17 & -0.02 &  0.11 &  0.27 &  0.11 &  0.47 \\ 
HR1903   & 26000 & 167 &  0.01 &  0.07 &  0.73 &  0.11 & -0.00 &  0.09 &  0.19 &  0.12 &  0.32 \\ 
HR1203   & 20700 & 160 & -0.01 &  0.12 &  0.80 &  0.18 &  0.01 &  0.11 &  0.23 &  0.11 &  0.56 \\ 
HR2827   & 13400 & 045 & -0.05 & -0.02 &  1.06 & -0.04 &  0.20 & -0.04 &  0.01 & -0.05 &  0.08 \\
HR1713   & 11200 & 218 &  0.04 &  0.06 &  1.27 &  0.23 &  0.05 &  0.09 &  0.19 &  0.17 &  0.50 \\ 
HR3975   &  9730 & 068 & -0.08 &  0.11 &  2.12 &  0.23 &  0.12 & -0.09 &  0.33 &  0.20 &  0.58 \\ 
HR7924   &  9080 & 225 & -0.05 &  0.00 &  1.40 &  0.20 & -0.02 &  0.03 & -0.14 &  0.08 &  0.08 \\ 
HR1865   &  7700 & 196 & -0.01 &  0.22 &  2.21 &  0.32 &  0.14 & -0.13 &  0.30 &  0.25 &  0.52 \\ 
HR1017   &  6640 & 248 &  0.08 &  0.25 &  1.88 &  0.43 &  0.06 & -0.02 &  0.53 &  0.41 &  1.05 \\ 
HR7796   &  6100 & 324 &  0.10 &  0.29 &  1.58 &  0.69 &  0.14 &  0.03 &  0.79 &  0.60 &  0.94 \\ 
HR8232   &  5510 & 290 &  0.09 &  0.37 &  1.28 &  0.67 &  0.17 &  0.03 &  1.18 &  0.67 &  1.19 \\ 
HR8752   &  5510 & 084 & -0.13 &  0.34 &  1.16 &  0.14 &  0.07 &  0.02 &  0.16 &  0.16 &  0.41 \\ 
HR7479   &  5333 & 096 &  0.15 &  0.13 &  1.13 &  0.70 &  0.10 &  0.00 &  0.92 &  0.41 &  0.69 \\ 
HR7479   &  5333 & 109 &  0.10 &  0.22 &  1.07 &  0.49 &  0.11 &  0.02 &  0.76 &  0.54 &  1.04 \\ 
HR7479   &  5333 & 182 &  0.13 &  0.12 &  1.08 &  0.56 &  0.05 &  0.07 &  0.94 &  0.49 &  1.06 \\ 
HR7479   &  5333 & 093 &  0.06 & -0.02 &  1.20 &  0.53 &  0.03 &  0.14 &  1.05 &  0.54 &  0.93 \\ 
HR7063   &  4902 & 260 &  0.17 &  0.12 &  0.94 &  0.84 &  0.14 &  0.25 &  1.66 &  0.76 &  1.06 \\ 
HR7314   &  4420 & 193 &  0.23 &  0.15 &  0.84 &  0.88 &  0.15 &  0.46 &  2.25 &  1.00 &  1.60 \\ 
HR7314   &  4420 & 257 &  0.15 &  0.17 &  0.67 &  0.59 &  0.10 &  0.37 &  1.91 &  0.86 &  1.36 \\ 
HR6713   &  4375 & 265 &  0.22 &  0.19 &  0.81 &  0.80 &  0.16 &  0.40 &  1.86 &  0.79 &  1.25 \\ 
HR8465   &  4295 & 308 &  0.13 &  0.28 &  0.68 &  0.70 &  0.19 &  0.63 &  2.35 &  1.02 &  1.79 \\ 
HR6498   &  4260 & 247 &  0.24 &  0.30 &  0.84 &  1.00 &  0.21 &  0.83 &  2.75 &  1.11 &  1.84 \\ 
HR603    &  4130 & 453 &  0.24 &  0.34 &  0.69 &  0.61 &  0.19 &  0.49 &  2.09 &  0.87 &  1.46 \\ 
HR8089   &  3990 & 202 &  0.21 &  0.45 &  0.50 &  0.67 &  0.14 &  0.68 &  2.26 &  0.94 &  1.59 \\ 
HR8079   &  3920 & 445 &  0.24 &  0.67 &  0.62 &  0.80 &  0.24 &  1.05 &  2.70 &  1.05 &  1.72 \\ 
HR2061   &  3550 & 327 &  0.54 &  1.08 &  0.36 &  1.39 &  0.23 &  1.46 &  3.92 &  0.89 &  2.22 \\ 
HR1155   &  3450 & 237 &  0.46 &  1.31 &  0.58 &  1.12 &  0.38 &  1.37 &  3.35 &  1.12 &  2.03 \\ 
HR921    &  2980 & 225 &  0.77 &  1.41 &  0.29 &  1.27 &  0.37 &  1.10 &  4.08 &  0.67 &  1.52 \\ 
HR7009   &  2925 & 292 &  0.63 &  1.43 &  0.79 &  1.55 &  0.54 &  1.79 &  4.68 &  1.26 &  2.21 \\ 
HR6406   &  2800 & 319 &  0.59 &  1.41 &  0.80 &  1.66 &  0.54 &  1.85 &  4.98 &  1.31 &  2.31 \\\tableline
\end{tabular} 
\label{ftshewsup}
\end{table}

\clearpage 
\vfill\eject

\begin{table}
\caption{
Equivalent Widths of Spectral Indices for Luminosity Class III Stars }
\begin{tabular}{llllllllllll} 
\tableline 
\multicolumn{1}{c}{Source} & 
\multicolumn{1}{c}{T$_{e}$ } & 
\multicolumn{1}{c}{SNR} & 
\multicolumn{1}{c}{Mg5844} & 
\multicolumn{1}{c}{OH5920} & 
\multicolumn{1}{c}{HI5950} & 
\multicolumn{1}{c}{Al5973} & 
\multicolumn{1}{c}{Si5993} & 
\multicolumn{1}{c}{CO6018} & 
\multicolumn{1}{c}{CO6170} & 
\multicolumn{1}{c}{Si6264} & 
\multicolumn{1}{c}{Mg6345} \\\tableline\tableline 
HR1899   & 32000 & 164 & -0.03 &  0.17 &  0.54 &  0.15 & -0.04 &  0.03 &  0.18 &  0.10 &  0.49 \\ 
HR1899   & 32000 & 182 & -0.04 &  0.20 &  0.62 &  0.11 &  0.05 & -0.03 &  0.07 & -0.00 &  0.07 \\ 
HR1552   & 20300 & 115 & -0.07 &  0.13 &  1.29 &  0.31 &  0.10 & -0.09 &  0.12 & -0.01 &  0.09 \\ 
HR5291   & 10100 & 159 & -0.07 &  0.19 &  2.65 &  0.78 &  0.02 &  0.05 &  0.28 &  0.11 &  0.41 \\ 
HR403    &  8100 & 202 & -0.02 &  0.39 &  2.79 &  0.77 &  0.09 & -0.13 &  0.25 &  0.08 &  0.54 \\ 
HR1412   &  7650 & 186 & -0.02 &  0.43 &  2.70 &  0.71 &  0.05 & -0.10 &  0.60 &  0.25 &  0.91 \\ 
HR4031   &  7150 & 193 &  0.02 &  0.29 &  2.30 &  0.54 &  0.12 & -0.09 &  0.25 &  0.13 &  0.48 \\ 
HR21     &  6870 & 245 &  0.03 &  0.32 &  1.99 &  0.55 &  0.04 & -0.11 &  0.20 &  0.22 &  0.82 \\ 
HR21     &  6870 & 179 &  0.01 &  0.25 &  2.21 &  0.63 &  0.13 & -0.14 &  0.46 &  0.24 &  0.69 \\ 
HR5017   &  6700 & 174 &  0.04 &  0.28 &  2.58 &  0.83 &  0.07 & -0.03 &  0.43 &  0.36 &  0.96 \\ 
HR2706   &  6470 & 086 &  0.01 &  0.39 &  2.08 &  0.47 &  0.16 & -0.10 &  0.13 &  0.20 &  0.64 \\ 
HR8905   &  6270 & 059 &  0.06 &  0.13 &  1.06 &  0.45 & -0.01 & -0.16 &  0.30 &  0.28 &  1.02 \\ 
HR4883   &  5910 & 280 &  0.15 &  0.20 &  1.07 &  0.52 &  0.09 &  0.04 &  0.90 &  0.43 &  1.02 \\ 
HR4716   &  5050 & 133 &  0.17 &  0.10 &  0.97 &  0.47 &  0.13 &  0.07 &  1.29 &  0.54 &  0.92 \\ 
HR7328   &  4885 & 281 &  0.19 &  0.20 &  0.78 &  0.57 &  0.11 &  0.11 &  1.33 &  0.67 &  1.35 \\ 
HR7949   &  4810 & 369 &  0.21 &  0.24 &  0.77 &  0.59 &  0.13 &  0.18 &  1.39 &  0.67 &  1.38 \\ 
HR8317   &  4710 & 191 &  0.29 &  0.14 &  0.63 &  0.58 &  0.08 &  0.26 &  1.79 &  0.83 &  1.64 \\ 
HR6299   &  4500 & 532 &  0.26 &  0.09 &  0.77 &  0.81 &  0.10 &  0.44 &  1.92 &  0.76 &  1.40 \\ 
HR165    &  4320 & 266 &  0.34 &  0.24 &  0.64 &  0.58 &  0.18 &  0.41 &  1.98 &  0.89 &  1.45 \\ 
HR6705   &  3990 & 298 &  0.39 &  0.60 &  0.77 &  1.12 &  0.26 &  0.96 &  3.05 &  1.12 &  1.75 \\ 
HR6705   &  3990 & 694 &  0.41 &  0.61 &  0.76 &  1.07 &  0.25 &  0.89 &  2.87 &  1.06 &  1.66 \\ 
HR152    &  3956 & 270 &  0.37 &  0.86 &  0.58 &  0.92 &  0.34 &  0.87 &  2.59 &  0.78 &  1.40 \\ 
HR4517   &  3780 & 673 &  0.66 &  1.06 &  0.33 &  1.07 &  0.21 &  0.89 &  3.14 &  0.70 &  1.93 \\ 
HR6242   &  3560 & 458 &  0.52 &  1.12 &  0.57 &  1.18 &  0.39 &  1.45 &  3.93 &  1.16 &  1.96 \\ 
HR7886   &  3250 & 574 &  0.61 &  1.54 &  0.89 &  1.71 &  0.64 &  2.25 &  5.69 &  1.41 &  2.42 \\\tableline
\end{tabular} 
\label{ftshewgn}
\end{table}

\clearpage 
\vfill\eject

\begin{table}
\caption{
Equivalent Widths of Spectral Indices for Luminosity Class IV Stars }
\begin{tabular}{llllllllllll} 
\tableline 
\multicolumn{1}{c}{Source} & 
\multicolumn{1}{c}{T$_{e}$ } & 
\multicolumn{1}{c}{SNR} & 
\multicolumn{1}{c}{Mg5844} & 
\multicolumn{1}{c}{OH5920} & 
\multicolumn{1}{c}{HI5950} & 
\multicolumn{1}{c}{Al5973} & 
\multicolumn{1}{c}{Si5993} & 
\multicolumn{1}{c}{CO6018} & 
\multicolumn{1}{c}{CO6170} & 
\multicolumn{1}{c}{Si6264} & 
\multicolumn{1}{c}{Mg6345} \\\tableline\tableline 
HR6588   & 19000 & 162 & -0.03 &  0.10 &  1.65 &  0.48 & -0.03 &  0.02 &  0.17 &  0.07 &  0.40 \\ 
HR4033   &  8810 & 146 & -0.00 &  0.33 &  2.62 &  0.73 &  0.06 &  0.02 &  0.27 &  0.13 &  0.35 \\ 
HR1351   &  7020 & 094 & -0.06 &  0.40 &  2.18 &  0.62 &  0.19 & -0.19 & -0.01 &  0.10 &  0.24 \\ 
HR5235   &  5930 & 341 &  0.16 &  0.12 &  1.21 &  0.83 &  0.11 &  0.03 &  1.30 &  0.64 &  1.14 \\ 
HR5235   &  5930 & 263 &  0.17 &  0.26 &  1.26 &  0.68 &  0.18 & -0.01 &  0.97 &  0.50 &  1.03 \\ 
HR5409   &  5830 & 158 &  0.16 &  0.16 &  1.02 &  0.51 &  0.12 & -0.03 &  0.91 &  0.44 &  0.95 \\ 
HR6623   &  5680 & 211 &  0.33 &  0.12 &  0.87 &  0.87 &  0.09 &  0.09 &  1.21 &  0.69 &  1.38 \\ 
HR995    &  5620 & 143 &  0.22 &  0.08 &  0.80 &  0.61 &  0.23 &  0.03 &  1.09 &  0.48 &  1.14 \\ 
HR7602   &  5430 & 056 &  0.46 &  0.15 &  0.71 &  0.86 &  0.01 &  0.17 &  1.40 &  0.57 &  1.49 \\ 
HR7957   &  5240 & 189 &  0.22 &  0.09 &  0.68 &  0.70 &  0.07 &  0.15 &  1.04 &  0.46 &  0.99 \\ 
HR7957   &  5240 & 192 &  0.09 &  0.18 &  0.56 &  0.52 & -0.01 &  0.09 &  0.75 &  0.40 &  1.17 \\ 
HR5901   &  5125 & 395 &  0.37 &  0.13 &  0.55 &  0.62 &  0.08 &  0.21 &  1.61 &  0.74 &  1.60 \\ 
HR5901   &  5125 & 153 &  0.37 &  0.09 &  0.75 &  0.93 &  0.15 &  0.30 &  1.54 &  0.71 &  1.13 \\ 
HR6014   &  5068 & 072 &  0.31 & -0.12 &  0.39 &  0.59 &  0.14 &  0.05 &  1.10 &  0.70 &  1.40 \\ 
HR6014   &  5068 & 133 &  0.42 &  0.11 &  0.58 &  0.55 &  0.18 &  0.09 &  1.48 &  0.65 &  1.24 \\\tableline  
\end{tabular} 
\label{ftshewsub}
\end{table}

\clearpage 
\vfill\eject

\begin{table}
\caption{
Equivalent Widths of Spectral Indices for Luminosity Class V Stars }
\begin{tabular}{llllllllllll} 
\tableline 
\multicolumn{1}{c}{Source} & 
\multicolumn{1}{c}{T$_{e}$ } & 
\multicolumn{1}{c}{SNR} & 
\multicolumn{1}{c}{Mg5844} & 
\multicolumn{1}{c}{OH5920} & 
\multicolumn{1}{c}{HI5950} & 
\multicolumn{1}{c}{Al5973} & 
\multicolumn{1}{c}{Si5993} & 
\multicolumn{1}{c}{CO6018} & 
\multicolumn{1}{c}{CO6170} & 
\multicolumn{1}{c}{Si6264} & 
\multicolumn{1}{c}{Mg6345} \\\tableline\tableline 
HR2456   & 38000 & 053 & -0.03 &  0.36 &  0.39 &  0.26 &  0.16 &  0.12 & -0.32 &  0.11 &  0.31 \\ 
HR2456   & 38000 & 073 & -0.12 &  0.18 &  0.48 &  0.10 &  0.06 & -0.15 &  0.15 &  0.03 &  0.10 \\ 
HR5191   & 19000 & 282 & -0.05 &  0.19 &  1.71 &  0.57 &  0.08 & -0.07 &  0.22 &  0.07 &  0.24 \\ 
HR3982   & 13000 & 241 & -0.08 &  0.13 &  2.22 &  0.47 &  0.04 & -0.09 &  0.28 &  0.03 &  0.15 \\ 
HR3982   & 13000 & 241 & -0.07 &  0.09 &  2.14 &  0.51 &  0.06 & -0.09 &  0.32 &  0.12 &  0.26 \\ 
HR7001   &  9480 & 678 & -0.05 &  0.36 &  2.83 &  1.06 &  0.03 & -0.03 &  0.41 &  0.15 &  0.52 \\ 
HR7001   &  9480 & 146 & -0.10 &  0.38 &  2.81 &  1.02 &  0.10 & -0.05 &  0.43 &  0.17 &  0.41 \\ 
HR2491   &  9145 & 111 & -0.07 &  0.36 &  2.71 &  0.83 &  0.16 & -0.14 &  0.37 & -0.08 &  0.10 \\ 
HR2491   &  9145 & 148 & -0.14 &  0.40 &  2.67 &  0.93 &  0.12 & -0.13 &  0.39 &  0.05 &  0.20 \\ 
HR4534   &  8593 & 205 & -0.04 &  0.45 &  2.67 &  1.01 &  0.11 & -0.12 &  0.30 &  0.07 &  0.34 \\ 
HR4357   &  8377 & 192 & -0.05 &  0.40 &  2.83 &  0.96 &  0.09 & -0.12 &  0.24 &  0.06 &  0.38 \\ 
HR4931   &  6750 & 119 &  0.04 &  0.24 &  1.47 &  0.62 & -0.01 & -0.06 &  0.09 &  0.09 &  0.86 \\ 
HR1279   &  6677 & 080 &  0.10 &  0.32 &  1.62 &  0.67 &  0.15 & -0.09 &  0.26 &  0.12 &  0.64 \\ 
HR2943   &  6530 & 327 &  0.04 &  0.33 &  1.60 &  0.57 &  0.15 & -0.14 &  0.54 &  0.27 &  0.74 \\ 
HR1538   &  6385 & 281 &  0.06 &  0.34 &  1.26 &  0.53 &  0.18 & -0.11 &  0.51 &  0.22 &  0.90 \\ 
HR4375   &  6085 & 254 &  0.26 &  0.15 &  0.63 &  0.56 &  0.15 & -0.06 &  0.68 &  0.37 &  1.09 \\ 
HR4983   &  5930 & 123 &  0.21 &  0.16 &  0.75 &  0.66 &  0.09 &  0.04 &  0.71 &  0.45 &  1.20 \\ 
HR4983   &  5930 & 173 &  0.19 &  0.21 &  0.90 &  0.52 &  0.19 & -0.06 &  0.87 &  0.47 &  0.96 \\ 
HR483    &  5855 & 059 &  0.24 &  0.02 &  0.68 &  0.29 &  0.04 & -0.17 &  0.18 &  0.37 &  1.09 \\ 
HR483    &  5855 & 079 &  0.24 &  0.19 &  0.90 &  0.46 &  0.19 & -0.11 &  0.67 &  0.48 &  1.30 \\ 
HR4374   &  5830 & 250 &  0.28 &  0.20 &  0.62 &  0.63 &  0.10 & -0.01 &  0.90 &  0.47 &  1.35 \\ 
HR5072   &  5740 & 197 &  0.26 &  0.15 &  0.74 &  0.53 &  0.16 & -0.04 &  0.99 &  0.53 &  1.28 \\ 
HR4496   &  5430 & 105 &  0.35 &  0.17 &  0.59 &  0.59 &  0.21 & -0.08 &  1.12 &  0.61 &  1.45 \\ 
HR4496   &  5430 & 090 &  0.31 &  0.16 &  0.39 &  0.50 &  0.13 & -0.04 &  0.69 &  0.43 &  1.08 \\ 
HR7462   &  5240 & 131 &  0.52 &  0.04 &  0.38 &  0.72 &  0.09 &  0.09 &  1.20 &  0.62 &  1.89 \\ 
HR7462   &  5240 & 095 &  0.51 &  0.07 &  0.37 &  0.69 &  0.13 &  0.06 &  1.16 &  0.72 &  2.05 \\ 
HR1084   &  5010 & 223 &  0.61 &  0.19 &  0.44 &  0.74 &  0.24 & -0.01 &  1.59 &  0.76 &  2.00 \\ 
HR8832   &  4785 & 111 &  0.88 &  0.14 &  0.29 &  1.06 &  0.17 &  0.18 &  1.88 &  0.90 &  2.50 \\ 
GL570A   &  4560 & 120 &  0.90 &  0.22 &  0.37 &  1.13 &  0.28 &  0.04 &  2.02 &  0.90 &  2.29 \\ 
HR8085   &  4340 & 125 &  0.65 &  0.18 &  0.01 &  1.15 &  0.11 &  0.15 &  1.59 &  0.60 &  2.38 \\ 
HR8086   &  4040 & 170 &  0.72 &  0.39 &  0.06 &  1.46 &  0.14 &  0.11 &  1.78 &  0.50 &  2.07 \\ 
HR8086   &  4040 & 181 &  0.72 &  0.32 &  0.07 &  1.48 &  0.15 &  0.12 &  1.86 &  0.52 &  2.18 \\ 
GL338A   &  3800 & 072 &  0.98 &  0.37 &  0.08 &  1.36 &  0.13 &  0.02 &  1.51 &  0.34 &  1.80 \\ 
GL526    &  3605 & 068 &  0.62 &  0.42 & -0.13 &  1.32 &  0.16 &  0.10 &  1.04 &  0.04 &  0.92 \\ 
GL411    &  3530 & 202 &  0.48 &  0.53 & -0.04 &  1.29 &  0.16 &  0.01 &  1.17 &  0.10 &  1.19 \\ 
GL411    &  3530 & 189 &  0.47 &  0.55 & -0.07 &  1.29 &  0.18 &  0.06 &  1.21 &  0.12 &  1.23 \\ 
GL725A   &  3380 & 083 &  0.52 &  0.57 & -0.10 &  1.15 &  0.06 &  0.03 &  0.93 & -0.00 &  0.99 \\ 
GL725A   &  3380 & 107 &  0.54 &  0.56 & -0.12 &  1.25 &  0.02 &  0.08 &  1.20 & -0.02 &  1.09 \\\tableline  
\end{tabular} 
\label{ftshewdw}
\end{table}

\end{tiny}

\clearpage 
\vfill\eject

%%%%%%%%%%%%%%%%%%%%%%%%%%%%%%%%%%%%%%%%%%%%%%%%%%%%%

\end{document}